\documentclass[12pt]{article}
\usepackage{amsmath}
\usepackage{amsfonts}
\usepackage{bbm}
\usepackage{graphicx,psfrag,epsf}
\usepackage{enumerate}
\usepackage{url} 

\newcommand{\blind}{0}

\usepackage[table]{xcolor}
\usepackage[section]{placeins}
\usepackage{xr-hyper}
\providecommand{\norm}[1]{\lVert#1\rVert}
\newcommand*\samethanks[1][\value{footnote}]{\footnotemark[#1]}

\emergencystretch=\maxdimen
\usepackage{hyperref}
\usepackage[style=apa, backend=biber, natbib=true, hyperref=true, uniquelist=false, sortcites]{biblatex}
\begin{document}

\def\spacingset#1{\renewcommand{\baselinestretch}%
{#1}\small\normalsize} \spacingset{1}


\if0\blind
{
  \title{\bf Functional clustering via multivariate clustering}
  \author{Belén Pulido \thanks{This research has been partially supported by Ministerio de Ciencia e Innovación, Gobierno de España, grant numbers PID2019-104901RB-I00 and PID2019-104681RB-I00 .}\hspace{.2cm}\\
    uc3m-Santander Big Data Institute, Universidad Carlos III de Madrid,\\
    Alba M. Franco-Pereira \samethanks\\
    Department of Statistics and O.R., Universidad Complutense de Madrid \\
    uc3m-Santander Big Data Institute, Universidad Carlos III de Madrid,\\
    and \\
    Rosa E. Lillo \samethanks\\
    Department of Statistics, Universidad Carlos III de Madrid\\
    uc3m-Santander Big Data Institute, Universidad Carlos III de Madrid}
  \maketitle
} \fi

\if1\blind
{
  \bigskip
  \bigskip
  \bigskip
  \begin{center}
    {\LARGE\bf Title}
\end{center}
  \medskip
} \fi

\bigskip
\begin{abstract}
Clustering techniques applied to multivariate data are a very useful tool in Statistics and have been fully studied in the literature. Nevertheless, these clustering methodologies are less well known when dealing with functional data. Our proposal consists of introducing a clustering procedure for
functional data using the very well known techniques for clustering multivariate data. The idea is to reduce a functional data problem to a multivariate data problem by applying the epigraph and the hypograph indexes to the original data and to its first and second derivatives. All the information given by the functional data is therefore transformed to the multivariate context, being sufficiently informative for the usual multivariate clustering techniques to be efficient. The performance of this new methodology is evaluated through a simulation study and it is also illustrated through real data sets.
\end{abstract}

\noindent%

\spacingset{1.45}
\section{Introduction}
\label{sec:intro}

Nowadays, in several fields of study, many of the data that are collected and analysed can be considered as functions $x_i(t)$, $i=1,...,n$, $t \in I$, where $I$ is an interval in $\mathbb{R}$. For example growth, weather variables, the evolution of the market... This is due to the fact that current technological development enables to analyse large volume of data in a short period of time. Functional Data Analysis (FDA) arises when this information is studied through the analysis of curves or functions. A complete overview of functional data analysis can be found in \citet{ramsay} and \citet{ferraty}, while some interesting reviews of functional data can be found in \citet{horvath}, \citet{hsing} and \citet{wang}.

The main problem when working with functional and multivariate data is that there does not exist a total order as in one dimension. Thus, a traditional challenge in FDA and in multivariate analysis is to provide an ordering within a sample of curves which enables the definition of order statistics such as ranks and $L$-statistics. In this sense, \citet{tukey} introduced the concept of statistical depth that provided a center-outward ordering for multivariate data. Some other definitions can be found in \citet{oja}, \citet{liu} and \citet{zuo}. This concept was extended to functional data appearing several definitions of functional depth. See, for example, \citet{vardi}, \citet{fraiman}, \citet{cuevas}, \citet{cuesta},  \citet{lop2009}, \citet{lop2011}, and \citet{sguera2014}. 

More recently, \citet{franc2011} proposed the epigraph and the hypograph indexes that, instead of dealing with the ``centrality" of a bunch of curves, allow to measure their ``extremality".
The combination of these indexes has been alredy exploited: \citet{arribas} proposed the outliergram for outliers detection,  \citet{mar2018} defined a boxplot for functional data and \citet{franc} contributed with a homogeneity test for functional data. The main idea of combining the epigraph and the hypograph indexes is to be able to summarize the information provided by a functional sample into a vector in $\mathbb{R}^2$ or in $\mathbb{R}^d$, $d \in \mathbb{N}$ if the derivatives of the functions are also considered. 

When studying high volume data, the necessity of classifying the data into groups without any extra information increases since they become easier to manipulate. Clustering is one of the most widely used techniques within unsupervised learning techniques, and that it has been fully studied for multivariate data. One of the most frequently used procedures, are the distance based techniques as hierarchical clustering (see \citet{Sibson}, \citet{Defays}, \citet{Sokal}, \citet{lance}, \citet{Ward} for different hierarchical clustering procedures) and k-means clustering (introduced by \citet{MacQueen}). Taking into consideration the fact that k-means is, probably, the mostly used clustering method in the literature, different variations of it have been introduced. See \citet{ben2001} and \citet{dhillon2004}.

Clustering functional data is a challenging problem since it involves working with an infinite dimensional space. Different approaches have been considered in the literature. In \citet{jacques2014} the functional clustering techniques are classified into four categories: raw data methods, which consist on considering the functional data set as a multivariate one and apply there the clustering techniques studied for multivariate data (\citet{boulle}); the filtering methods, that firstly apply a basis to the functional data and then use clustering techniques to the obtained data (\citet{abraham2003}, \citet{rossi2004}, \citet{peng2008}, \citet{kayano2010}); adaptive methods, where dimensionality reduction and clustering are performed at the same time (\citet{james2003}, \citet{jacques2013}, \citet{giacofci2013wavelet}, \citet{traore2019}); and distance-based methods, which apply a clustering technique based on distances with a specific distance for functional data (\citet{tarpey2003}, \citet{ieva2011}, \citet{martino}). Recent works which perform different strategies for clustering functional data are \citet{zambom2019} that propose a new method applying k-means, assigning each element to a cluster or another based on a combination of an hypothesis test of parallelism and a test for equality of means, and \citet{schmutz2020} that presents a new strategy for clustering functional data based on applying model based techniques when a principal component analysis is previously performed.  

In this paper we propose a new techique for clustering functional data based on the used of the epigraph and the hypograph indexes and their modified versions. The idea is to transform a functional data problem into a multivariate one and then use the very well known techniques for multivariate clustering.
 
The paper is organized as follows. In Section \ref{sec:ind}, the epigraph and the hypograph indexes are introduced and the methodology for clustering functional data sets based on these indexes is explained in Section \ref{sec:met}. Some other techniques for clustering functional data are presented in Section \ref{sec:benchmark}. In Section \ref{sec:sim} we present the results of an extensive simulation study in which our proposed methodology is compared to the existing procedures and, in Section \ref{sec:realdata}, we illustrate its applicability through some real data sets. In section \ref{sec:nclus} we present an important question, which is the election of the number of clusters prior the application of any clustering methodology, and in Section \ref{sec:conc}, a brief discussion and the main conclusions of the paper are reflected.
 
\section{Preliminaries: The epigraph and the hypograph indexes}
\label{sec:ind}

Let $C(I)$ be the space of continuous functions defined on a compact interval $I$. Let consider a stochastic process $X$ with sample paths in $C(I)$ and distribution $F_{X}$. The graph of a function $x$ in $C(I)$ is $G(x) = \{(t,x(t)),t \in I \}.$ Then, the epigraph (epi) and the hypograph (hyp) of $x$ are defined as follows:
$$epi(x)=\{(t,y) \in I \times \mathbb{R} : y \geq x(t)\},$$
$$hyp(x)=\{(t,y) \in I \times \mathbb{R} : y \leq x(t)\}.$$

Taking into account the information that can be obtained from these graphs, \citet{franc2011} defined two indexes based on these two concepts. Given a sample of curves $\{x_1(t),...,x_n(t)\}$, the epigraph index of a curve $x$ ($EI_n(x)$) is defined as one minus the proportion of curves in the sample that are totally included in its epigraph. Analogously, the hypograph index of $x$ ($HI_n(x)$) is the proportion of curves totally include in the hypograph of $x$.

$$EI_n(x)=1-\frac{ \sum_{i=1}^n{I(\{G(x_i)\subseteq epi(x)\})}}{n}=1-\frac{\sum_{i=1}^n{I(\{x_i(t)\geq x(t), t \in I\})}}{n},$$
$$HI_n(x)=\frac{ \sum_{i=1}^n{I(\{G(x_i)\subseteq hyp(x)\})}}{n}=\frac{ \sum_{i=1}^n{I(\{x_i(t)\leq x(t), t \in I\})}}{n}.$$
Their population versions are given by: 
$$EI(x,F_X)\equiv EI(x) = P(G(X) \subseteq epi(x))=1-P(X(t)\geq x(t),t\in I),$$ $$HI(x,F_X)\equiv EI(x) = P(G(X) \subseteq hyp(x))=1-P(X(t)\leq x(t),t\in I).$$ 

\citet{franc2011} argued that when the curves in the sample are extremely irregular, having lots of intersections, the modified versions of these indexes are more convenient. If $I$ is considered as a time interval, the modified epigraph index of $x$ ($MEI_n(x)$) can be defined as one minus the proportion of time that the curves of the sample are in the epigraph of a given curve, i.e., the proportion of time that the curves of the sample are above it. Analogously, the generalized hypograph index of $x$ ($MHI_n(x)$) can be considered as the proportion of time the curves in the sample are below a given curve.
\begin{equation}\label{MEI}
    MEI_n(x)=1- \sum_{i=1}^n{ \frac{ \lambda (\{G(x_i)\subseteq epi(x)\})}{n \lambda(I)}},
\end{equation}
\begin{equation}\label{MHI}
   MHI_n(x)= \sum_{i=1}^n{ \frac{ \lambda (\{G(x_i)\subseteq hyp(x)\})}{n \lambda(I)}},
\end{equation}
 where $\lambda$ stands for the Lebesgue's measure on $\mathbb{R}$.

Note that, since the graph of any curve $x$ is contained in its epigraph and its hypograph, this relation holds: $$\lambda(\{ G(x) \subseteq epi(x) \}) = \lambda(I) = \lambda(\{G(x) \subseteq hyp(x)\}).$$

Applying this condition to (\ref{MEI}) and (\ref{MHI}), we obtain
\begin{equation}\label{MEI1}
     MEI_n(x)=1- \left(\sum_{\substack{i=1 \\ x_i \neq x}}^n{ \frac{ \lambda (\{G(x_i)\subseteq epi(x)\})}{n \lambda(I)}} + \frac{1}{n}\right),
\end{equation}
and
\begin{equation}\label{MHI1}
     MHI_n(x)=\sum_{\substack{i=1 \\ x_i \neq x}}^n{ \frac{ \lambda (\{G(x_i)\subseteq hyp(x)\})}{n \lambda(I)}} + \frac{1}{n}.
\end{equation}

Moreover, if $x\neq x_i$, for all $i=1,...,n$, then $$\lambda(\{G(x_i)\subseteq hyp(x)\}) +\lambda(\{G(x_i)\subseteq epi(x)\}) = \lambda(I), $$ and applying this into (\ref{MHI1}) we can write: $$MHI_n(x) = 1-\frac{1}{n}- \sum_{\substack{i=1 \\ x_i \neq x}}^n{ \frac{ \lambda (\{G(x_i)\subseteq epi(x)\})}{n \lambda(I)}} + \frac{1}{n} \stackrel{(\ref{MEI1})}{=} MEI_n(x)+\frac{1}{n}. $$

Finally, we obtain the following relation between the two modified versions of the epigraph and the hypograph indexes, leading to conclude that both are linearly dependent: $$MHI_n(x)-MEI_n(x)=\frac{1}{n}.$$

Note that this equality does not hold in \citet{franc} because of the way in which the data is considered in the homogeneity test differs from the perspective given in this paper.

\section{Clustering functional data through the epigraph and the hypograph indexes}
\label{sec:met}
\begin{figure}[htb]
    \centering
    \includegraphics[scale=0.9]{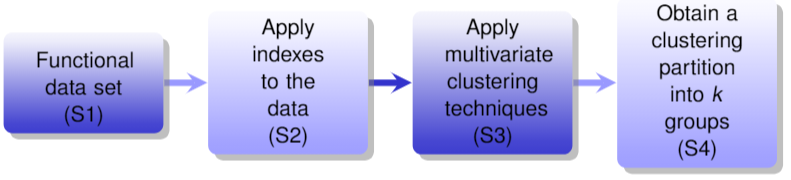}
    \caption{Scheme of the proposed methodology for clustering functional data.}
    \label{fig2}
\end{figure}

The four steps of the proposed methodology for clustering functional data are illustrated in Figure \ref{fig2}.
\vspace{-0.3cm}

Step 1 (S1) consists of smoothing the data. This is recommended since the amount of the data upon which the process is based precludes abrupt changes in value. For this reason, it is common to smooth the sample curves when working with functions. We have used a cubic B-spline basis, but any other functional basis could have been used. After the data set is transformed, the second step (S2) is to apply the epigraph and the hypograph indexes (and their generalized versions) to the basis transformed data, as well as to their derivatives, obtaining a multivariate data set. Then, the combination of data that is the most informative is considered and a multivariate clustering technique is applied in the third step of the process (S3). Finally, the fourth step (S4) consists of obtaining a final clustering partition in the previously set number of groups.

\begin{figure}[htb]
    \centering
    \includegraphics[scale=0.85]{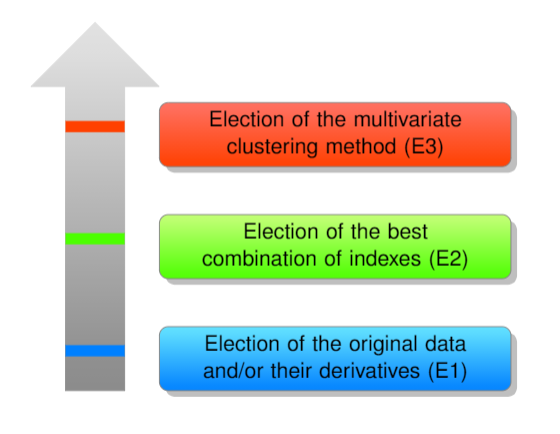}
    \caption{The three elections to be made during the proposed procedure.}
    \label{fig1}
\end{figure}

Along the procedure, it is necessary to make three elections as represented in Figure \ref{fig1}. These elections will be explained below.

\textbf{Election of the data and/or their derivatives (E1).}  Once dealing with smoothed functions, it is possible to consider their derivatives. When the indexes are applied to the data (S2), we move from a functional data problem into a multivariate one. Thus, we loose some ``information". Applying the indexes, not only to the original functions but also to their first and second derivatives, allows us to take advantage of the shape of the functions in the sample. When one poses a problem of discrimination with functions, it is clear that the shape of these functions must be taken into account, and the epigraph and the hypograph indexes have the property of reflecting the shape of the data, as it was shown in \citet{mar2018}. From now on, the term data we will be used to refer to the original functional sample together with the corresponding first and second derivatives.

\textbf{Election of the best combination of indexes (E2).} As explained in Section \ref{sec:ind}, the modified epigraph and hypograph indexes are linearly dependent. Because of that, MEI will be discarded when obtaining the multivariate data set through the indexes since it will not provide ``extra information".
    
First, the simple epigraph and hypograph indexes are considered together for each set of data, and then those tuples are combined. We proceed analogously with the modified epigraph index. All these sets can be studied together, obtaining eighteen different sets of variables. These combinations sets are represented as (b).(c) where (b) stands for the data, where `$\_$' represents the original data, `d' first derivatives and `d2' second derivatives, and (c) represent the indices. Some examples of the vectors represented by (b).(c) are shown in Table \ref{table_ex}:
\begin{table}[htbp]
\begin{centering}
    \begin{tabular}{|m{6cm}|m{6cm}|}
    \hline
    Representation & Explanation \\ \hline
       $\_$.EIHI = (EI, HI)  & The epigraph and the hypograph index on the original data. \\
       \hline
       dd2.MEI = (dMEI, d2MEI)  & The generalized epigraph index applied to the first and second derivatives of the data. \\
       \hline
        $\_$dd2.EIHIMEI =  (EI, HI, MEI, dEI, dHI, dMEI, d2EI, d2HI, d2MEI) & The epigraph, the hypograph and the generalized version of the epigraph applied to all the data.\\
        \hline
    \end{tabular}
    \caption{Representation and explanation of the combinations of data and indexes.}
\label{table_ex}
\end{centering}
\end{table}

When curves are extremely irregular, the epigraph and the hypograph indexes take values very close to 1 and 0, respectively. This fact causes that the indexes lose discriminatory capacity to differentiate between clusters and also induce computational problems. That is, singular or near-singular matrix is often referred to as ``ill-conditioned" matrix because it delivers problems in many statistical data analysis. To solve this issue, a condition regarding the variability of the indexes is included for each combination of indexes to be considered in the clustering process. This condition is $$|det(Var(Y))| > 10^{-5},$$ where $Y$ is a matrix of shape \textit{number of curves $\times$ number of indexes}. 

 \textbf{Election of the multivariate clustering method (E3).} We have considered the following: hierarchical clustering techniques using different criteria for calculating similarity between clusters (single linkage, complete linkage, average linkage, centroid linkage), and Ward's method. Besides, we have also considered k-means method and its different versions using a feature space induced by a kernel, such as kernel k-means (kkmeans), spectral clustering (spc) and  support vector clustering (svc).

For hierarchical methods, the Euclidean distance has been considered. On the other hand, when implementing $k$-means, Euclidean and Mahalanobis distances have been used. In order to apply the Mahalanobis distance, the data is rescaled using the Cholesky decomposition of the variance matrix before running $k$-means with the Euclidean distance (see \citet{km_mahal}). 
Moreover, when the method use a kernel space we have applied three type of kernels: gaussian, polynomial and linear.

In summary, a clustering partition is obtained once we choose the data to be considered (E1), the combination of indexes to apply (E2) and the multivariate clustering method (E3).

Once we have described the methodology, it is necessary to apply an external validation criterion in order to evaluate how it works. We have considered three different multivariate clustering evaluation criteria to rank the results of the algorithms. Purity, F-measure and Rand Index (RI), that are fully explained in \citet{indexes} and \citet{rendon}.

Purity of a cluster measures the fraction of the cluster size with the most repeated value. Because of that, purity scores positively the fact that there is a unique group in the clustering partition. On the other hand, F-measure penalizes this fact by focusing on the overlapping between obtained and real classifications. A small value indicates that the clustering partition only has one class. Rand Index applies a different approach. Instead of counting single elements, RI counts pairs of elements that are correctly or incorrectly classified. All these indexes provide values in $[0,1]$ and verify that the higher its value is, the better the classification.

\section{The benchmark for clustering functional data}
\label{sec:benchmark}
There are several studies concerning clustering functional data, that show the interest this topic arouses. Here, we have considered two recent works: The distance based k-means procedure (functional k-means) appearing in \citet{martino} and the test based k-means from \citet{zambom2019}. 

In \citet{martino}, a clustering procedure based on k-means clustering with the generalized Mahalanobis distance, $d_{\rho}$, previously defined in \citet{ghiglietti}, where the values of $\rho$ have to be set in advance is proposed. 

Let $X_i$, $X_j$, $i,j \in \mathbb{N}$ be two realizations of a stochastic process $X$. Then, $$d_{\rho}(X_i,X_j) = \displaystyle \sqrt{\sum_{l=1}^\infty d_{M,l}^2(X_i,X_j)h_l(\rho)},$$ where $d_{M,l}$ stands for the contribution of the Mahalanobis distance to this generalized one and which is defined as $$d_{M,l}(X_i,X_j)= \frac{\langle X_i-X_j, \varphi_l\rangle^2}{\lambda_l}. $$ Here, $\varphi_l$ and $\lambda_l$ stand for the eigenfunctions and eigenvalues of the covariance kernel, and $h_l(\rho)$ is a sequence of functions for a given value $\rho$ that makes it possible that the generalized Mahalanobis distance verify the conditions for being a function. These functions are defined as $$h_l(\rho)=\int_0^\infty \lambda_l\exp(-\lambda_l c) \ g_\rho(c) \ dc,$$ such that the function $g_\rho(s)$ is considered to deal with the situation in which \\ $d_{M,l}(X_i,X_j)\exp(-\lambda_ls)$ is a finite function but not integrable for every $s$.

The values of $\rho$ considered in their paper, and that we have considered in our simulation study for comparison, are, $\rho_1=0.001$, $\rho_2=0.02$, $\rho_3=1$, $\rho_4=100$ and $\rho_5=10^8$. The data is smoothed with a B-spline basis before applying k-means. Besides, they compare their procedure performance with the results of $k$-means applied with the truncated Mahalanobis distance, $d_K$, choosing a fixed number $K$ of principal components defined as $$d_K(X_i,X_j)=\displaystyle \sqrt{\sum_{l=1}^K \widehat d_{M,l}^2(X_i,X_j)}, $$ where $\widehat d_{M,l}^2$ stands for the empirical version of $ d_{M,l}^2$. They also compare their results to those obtained from $k$-means applied with the $L^2$ distance defined as follows (see \citet{horvathe}) $$d_{L^2}(X_i,X_j)=\norm{X_i-X_j},$$ being $I$ a compact interval. They have developed the `gmfd' R-package that is associated to the cited paper, and which will be used to compare the two methodologies.

\citet{zambom2019} propose a methodology based on an hypothesis test applying k-means, where they initialize the clusters centers in four different ways, choosing them at random, choosing one iteration of k-means, one iteration of a hierarchical method (Ward's method with Euclidean distance) or one iteration of k-means$++$ (\citet{vas}). 

At each step of the k-means algorithm, the allocation of each curve to a cluster is based on an hypothesis test performed as the combination of two test statistics. 

Let $X_i(t_i^j)$, $i = 1,...,n$, $j = 1...,, r_i$ the realizations of a stochastic process $X$, and $\{c^p(t_i^j), p = 1,...,K, \ i = 1,...,n, \ j = 1...,, r_i\}$ the set of estimated values of the cluster centers at grid point $t_i^j$, where $K$ is the previously selected number of clusters. The first statistic measures the proximity between the curve and the cluster centers by looking for parallelism. The residuals are computed as $\xi_i^{jp} = X_i(t_i^j)-c^p(t_i^j)$, and they consider $(\xi_i^{jp}, t_i^j)$ as data from a one-way ANOVA design with $\xi_i^{jp}$ being the observation at ``level" $t_i^j$. As two or more observations per factor level are required, each cell $t_i^j$ is augmented by including the $\xi_i^{jp}$ corresponding to $(m-1)$, $m$ odd, nearest grid points on either side, i.e. $W_i^j = \{s, \lvert t_i^j - t_i^s \rvert \leq \frac{m-1}{2r_i}\}$. The window size $m$ determines the number of neighbors included in each augmented ANOVA level. Then, they compute the test statistic for parallelism as the absolute value of the standardized statistic $$T(\xi_i^p) = \displaystyle\left\lvert \frac{\sqrt{n}(MST(\xi_i^p) - MSE(\xi_i^p)}{\widehat{\tau_{ip}}\sqrt{2m(2m-1)/3(m-1)}}\right\rvert,$$ where $\xi_i^p = (\xi_i^{1p},...,\xi_i^{rp})$ and $\widehat{\tau_{ip}}^2=\frac{1}{4(n-3)}\sum_{s=2}^{r_i-2}(\xi_i^{sp}- \xi_i^{(s-1)p})^2(\xi_i^{(s+2)l}- \xi_i^{(s+1)l})^2$.

The second statistic tests for equality of means using the $t$ test statistic for differences in averages. They consider $$W(X_i,c^p)=\displaystyle\left\lvert\frac{\frac{1}{r_i}\sum_{j=1}^{r_i}X_i(t_i^j) - \frac{1}{r}\sum_{j=1}^{r_i}c^p(t_i^j)}{\sqrt{\frac{\widehat{V}(X_i)+\widehat{V}(c^p)}{r_i}}}\right\rvert,$$ where $X_i = (X_i(t_i^j),...,X_i(t_i^{r_i}))$, $c^p = (c^p(t_i^1),...,c^p(t_i^{r_i}))$ and $\widehat{V}$ is an unbiased estimator of their variance.

Finally, they propose allocating the $i$th curve to the $p$th cluster center by combining the two previously defined statistics in the following way:

$$\Psi_{ip}=\begin{cases}
W(X_i, c^p), \  \text{if $\displaystyle\left(\sum_{p=1}^K I(T(\xi_i^p)<\gamma)\geq 2\right)$ and $\displaystyle\left(\sum_{p=1}^K I(W(X_i,c^p)<\gamma) \leq 1\right)$},\\[18pt]
T(\xi_i^p), \  \text{if $\displaystyle\left(\sum_{p=1}^K I(W(X_i,c^p)<\gamma) \geq 2 \right)$ and $\displaystyle\left(\sum_{p=1}^K I(T(\xi_i^p)<\gamma) \leq 1 \right)$}, \\[18pt]
\displaystyle\left(\frac{1}{2}\frac{T(\xi_i^p)}{max\{T(\xi_i^p)\}} + \frac{1}{2} \frac{W(X_i,c^p)}{max\{W(X_i,c^p)\}}\right), \  \text{otherwise,}
\end{cases}$$ where $I(\cdot)$ denotes the indicator function and $\gamma$ determines the rejection tail threshold for the test statistics.

Selecting the number of clusters prior the application of any clustering techniques is an interesting topic that is still open. These two recent techniques have been proposed fixing the number of clusters in advanced. Choosing the number of clusters suppose a limitation that have been studied, and which is now a work in progress to be further developed. Moreover, these techniques are computationally expensive, which suppose another important limitation. In order to find a competitive and faster alternative, in the following section we compare our methodology to those two.

\section{Simulation study}
\label{sec:sim}
We have carried out a simulation study in order to evaluate our methodology and to compare it with the functional k-means and the test based k-means strategies.

Each simulated scenario consists of a previously known number of groups proceeding from different processes. Each scenario is simulated 100 times for each of the three methodologies. In each iteration, we compute the average of the three validation criteria: Purity, F-measure, Rand Index, and the mean execution time of all the methods. We have divided this section into two depending on the previously known number of clusters.  The code necessary to develop the simulation is available in \url{https://github.com/bpulidob/Functional-clustering-via-multivariate-clustering}.

\FloatBarrier
\subsection{Simulation study A: Two clusters}
Two different simulation group of scenarios will be studied in this section. The first one consist of eight different scenarios that have been previously considered in \citet{flores} and \citet{franc}. The second one consist of two scenarios studied in \citet{martino}.

First we describe how we simulate the data of the first group of scenarios: Consider eight functional samples defined in $[0,1]$, which have continuous trajectories in such interval and which are the realizations of a stochastic process $X$. Each curve has 30 equidistant observations in the interval $[0,1]$. We generate 100 functions: 50 from Model 1 and 50 from Model $i$, $i=2,...,9$, obtaining eight different functional data sets.  

\begin{itemize}
\item[]\textbf{Model 1.} This is the set of functions which is considered in all the eight data sets for generating the first 50 functions. It is generated by a Gaussian process $$X_1(t)=E_1(t)+e(t),$$ where $E_1(t)=E_1(X(t))=30t^{\frac{3}{2}}(1-t)$ is the mean function and $e(t)$ is a centered Gaussian process with covariance matrix $Cov(e_i,e_j)=0.3 \ \exp (-\frac{|t_i-t_j|}{0.3})$.

The rest of models are obtained from the first one by perturbing the generation process. 

The first three models contain changes in the mean, while the covariance matrix does not change. Changes in the mean are presented in increasing order from Model 2 to Model 4. 

\item[]\textbf{Model 2.} $X_2(t)=30t^{\frac{3}{2}}(1-t)+0.5+e(t).$

\item[]\textbf{Model 3.} $X_3(t)=30t^{\frac{3}{2}}(1-t)+0.75+e(t).$

\item[]\textbf{Model 4.} $X_4(t)=30t^{\frac{3}{2}}(1-t)+1+e(t).$

The next two samples are obtained by multiplying the covariance matrix by a constant.

\item[]\textbf{Model 5.} $X_5(t)=30t^{\frac{3}{2}}(1-t)+2 \ e(t).$

\item[]\textbf{Model 6.} $X_6(t)=30t^{\frac{3}{2}}(1-t)+0.25 \ e(t).$

\item[]\textbf{Model 7.}  This set is obtained from adding to $E_1(t)$ a centered Gaussian process $h(t)$ whose covariance matrix is given by $Cov(e_i,e_j)=0.5 \  \exp (-\frac{|t_i-t_j|}{0.2})$. So, in this case $X_7(t)=30t^{\frac{3}{2}}(1-t)+ \ h(t).$

The next two samples are obtained by a different mean function.

\item[]\textbf{Model 8.} $X_8(t)=30t{(1-t)}^2+ h(t).$
\item[]\textbf{Model 9.} $X_9(t)=30t{(1-t)}^2+ e(t).$
\end{itemize}

From now on, the eight resulting data sets will be referred as
scenarios, where \textit{S 1-2} refers to the combination of Model 1 and 2, \textit{S 1-3} obtained from combining Model 1 and 3, and so on.

We smooth the data using a cubic B-spline basis in order to remove noise and to use the derivatives of the data (S1 in Figure \ref{fig2}). Then, we simulate each of the scenarios 100 times and apply, each time, our whole clustering strategy (S1-S4 in Figure \ref{fig2}). The three elections of the process (E1, E2, E3 in Figure \ref{fig1}) are carried out for each simulated data set. The mean Purity, F-measure and Rand index (RI), that are used as a criterion to choose the best model, as well as the execution time (ET) for the scenario \textit{S 1-4} are shown in Table \ref{tables3}. The rest of the tables are deferred to the Supplementary Material. In these tables, each row represents a description of the process carried out for the 100 realizations, denoted by (a).(b).(c) where $(a)$ represents the name of the considered strategy: a hierarchical method, k-means, support vector clustering, kernel k-means or spectral clustering, and $(b).(c)$ represents the elections of the data and indices, as represented in Table \ref{table_ex}, where (b) is the name of the employed data and (c) the applied indexes on the corresponding data.

\begin{figure}[ht]
    \centering
    \includegraphics[scale=0.55]{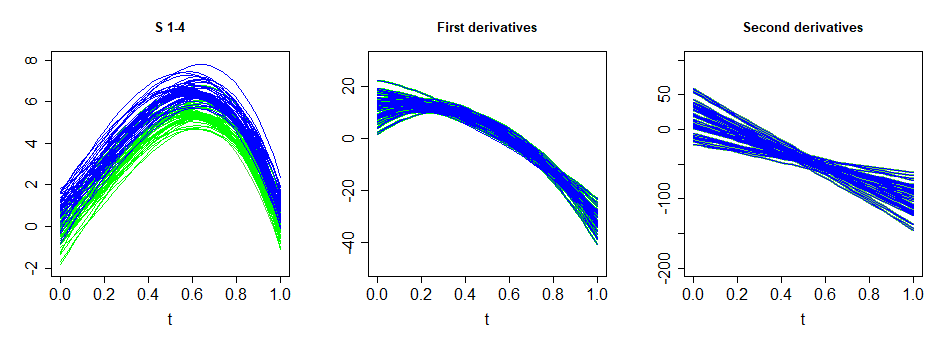}
    \caption{A sample generated from \textit{S 1-4}. Original data, first and second derivatives curves.}
    \label{figsc3f}
\end{figure}

\begin{figure}[ht]
    \centering
    \includegraphics[scale=0.55]{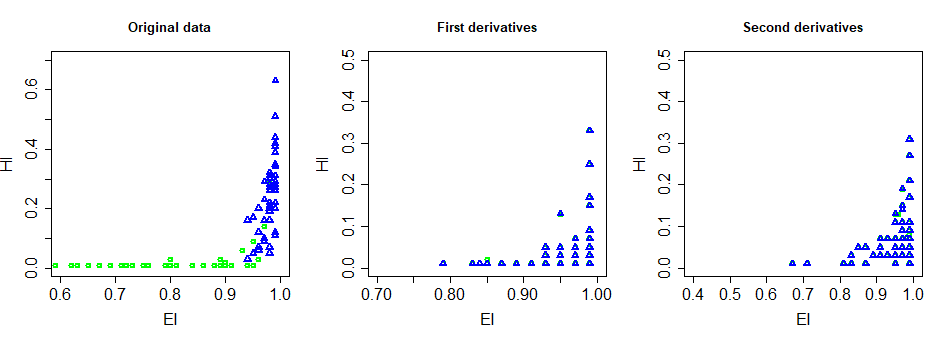}
    \caption{Scatter plots of the epigraph index (EI) and the hypograph index (HI) of the original data simulated from Model 1 and 4 (left panel), first derivatives (center panel) and second derivatives (right panel).}
    \label{figsc3ind}
\end{figure}

\begin{figure}[ht]
    \centering
    \includegraphics[scale=0.55]{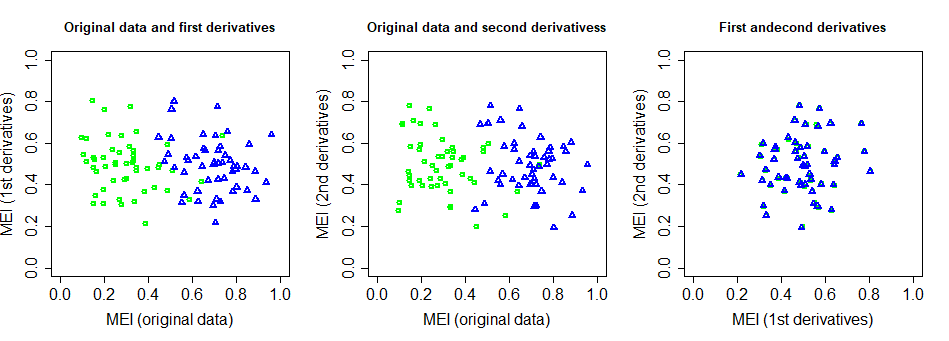}
    \caption{A sample generated from \textit{S 1-4}. Scatter plots of different combinations of MEI.  Original data and first derivatives (left panel), original data and second derivatives (center panel) and first and second derivatives (right panel).}
    \label{figsc3Mind}
\end{figure}

The smoothed functions and the first and second derivatives of data generated from \textit{S 1-4} are shown in Figure \ref{figsc3f}. It is clear that, in this case, the original data better discriminate the two clusters. When the epigraph and the hypograph indexes (Figure \ref{figsc3ind}), and the modified epigraph index of different combinations of variables (Figure \ref{figsc3Mind}) are applied, this is also noticeable since the combinations which better distinguish between the two clusters are those including the original data. In these figures only the combinations of two variables are shown. However, combinations up to nine variables are possible.

An important fact is that first and second derivatives of the two groups are the same because of the nature of the functions. Thus, in this scenario the first and second derivatives are not important for classification when they are considered together. Besides, turn out to be more informative those combinations including the original data (Table \ref{tables3}), where the best result, with a RI of 0.860, is obtained from applying support vector clustering on the epigraph and hypograph index of the original data (svc.$\_$.EIHI).

\begin{table}[htbp]
\vspace{-1cm}
\hspace*{-2cm}
\centering
\parbox{10cm}{

\footnotesize
\scalebox{0.9}{
\begin{tabular}{lcccc}
  \hline
 & Purity & Fmeasure & RI & Time \\ 
  \hline \rowcolor{green!10}
 svc.$\_$.EIHI & 0.924 & 0.859 & 0.860 & 0.00237 \\ \rowcolor{orange!20}
   svc.$\_$.EIHI & 0.924 & 0.859 & 0.860 & 0.00260 \\\rowcolor{cyan!10} 
  kkmeans.$\_$.EIHI & 0.924 & 0.859 & 0.860 & 0.00600 \\ \rowcolor{gray!10}
   kmeans.$\_$.EIHI & 0.923 & 0.857 & 0.858 & 0.00121 \\ \rowcolor{magenta!10}
  kmeans.$\_$.EIHI & 0.923 & 0.857 & 0.858 & 0.00125 \\ \rowcolor{gray!10}
   kmeans.$\_$d.MEI & 0.923 & 0.857 & 0.858 & 0.00126 \\ \rowcolor{magenta!10}
   kmeans.$\_$d.MEI & 0.923 & 0.857 & 0.858 & 0.00123 \\ \rowcolor{orange!20}
   svc.$\_$d.MEI & 0.923 & 0.857 & 0.858 & 0.00230 \\ \rowcolor{cyan!10} 
   kkmeans.$\_$d.MEI & 0.922 & 0.855 & 0.857 & 0.00529 \\ \rowcolor{green!10}
   svc.$\_$d.MEI &  0.922 & 0.855 & 0.857 & 0.00225 \\ \rowcolor{cyan!10}
   kkmeans.$\_$d2.MEI & 0.920 & 0.851 & 0.852 & 0.00536 \\ \rowcolor{gray!10}
   kmeans.$\_$d2.MEI & 0.919 & 0.851 & 0.852 & 0.00111 \\ \rowcolor{magenta!10}
   kmeans.$\_$d2.MEI & 0.919 & 0.851 & 0.852 & 0.00117 \\ \rowcolor{orange!20}
   svc.$\_$d2.MEI & 0.919 & 0.850 & 0.851 & 0.00271 \\ \rowcolor{orange!20}
   svc$\_$dd2.MEI & 0.919 & 0.850 & 0.851 & 0.002 \\ \rowcolor{green!10} 
   svc.$\_$d2.MEI & 0.919 & 0.850 & 0.851 & 0.00225 \\ \rowcolor{gray!10}
   kmeans$\_$dd2.MEI & 0.919 & 0.849 & 0.850 & 0.00126 \\ \rowcolor{magenta!10}
  kmeans$\_$dd2.MEI & 0.919 & 0.849 & 0.850 & 0.00136 \\ \rowcolor{magenta!10}
   svc$\_$dd2.MEI & 0.918 & 0.848 & 0.849 & 0.00291 \\ \rowcolor{cyan!10}
   kkmeans$\_$dd2.MEI & 0.912 & 0.843 & 0.845 & 0.00575 \\ \rowcolor{yellow!20}
   kkmeans$\_$dd2.MEI & 0.899 & 0.826 & 0.827 & 0.00861 \\ \rowcolor{gray!10}
   average.$\_$d2.MEI & 0.901 & 0.824 & 0.822 & 0.00018 \\ \rowcolor{gray!10}
   average.$\_$d.MEI & 0.900 & 0.822 & 0.821  & 0.00071 \\ \rowcolor{gray!10}
   ward.D2.$\_$d2.MEI & 0.896 & 0.816 & 0.814 & 0.00016 \\ \rowcolor{gray!10}
  ward.D2.$\_$d.MEI & 0.894 & 0.814 & 0.812 & 0.00016 \\ \rowcolor{gray!10}
  ward.D2$\_$dd2.MEI & 0.892 & 0.809 & 0.808 & 0.00018 \\ \rowcolor{gray!10}
   average$\_$dd2.MEI & 0.891 & 0.809 & 0.807 & 0.00022 \\ \rowcolor{gray!10}
   ward.D2.$\_$.EIHI & 0.890 & 0.811 & 0.807 & 0.00019 \\ \rowcolor{gray!10}
   complete$\_$dd2.MEI & 0.883 & 0.799 & 0.797 & 0.00071 \\ \rowcolor{gray!10}
   complete.$\_$d.MEI &  0.882 & 0.796 & 0.793 & 0.00015 \\ \rowcolor{gray!10}
   complete.$\_$d2.MEI & 0.878 & 0.793 & 0.790 & 0.00018 \\ \rowcolor{gray!10}
   centroid.$\_$d.MEI & 0.864 & 0.805 & 0.790 & 0.00018 \\ \rowcolor{yellow!20}
   kkmeans.$\_$d2.MEI & 0.865 & 0.786 & 0.785 & 0.00873 \\ \rowcolor{gray!10}
   average.$\_$.EIHI & 0.862 & 0.800 & 0.784 & 0.00015 \\ 
   spc.$\_$d2.MEI &  0.833 & 0.789 & 0.761 & 0.02677 \\ 
   spc.$\_$d.MEI & 0.829 & 0.789 & 0.759 & 0.02592 \\ 
         \end{tabular}
   }}
\parbox{9cm}{
\footnotesize
\scalebox{0.9}{
\begin{tabular}[t]{lcccc}
  \hline
 & Purity & Fmeasure & RI & Time \\ 
  \hline \rowcolor{yellow!20}
   kkmeans.$\_$d.MEI & 0.837 & 0.757 & 0.755 & 0.00865 \\ 
   spc$\_$dd2.MEI & 0.820 & 0.786 & 0.753 & 0.02604 \\
   spc.$\_$.EIHI & 0.816 & 0.791 & 0.752 & 0.02790 \\\rowcolor{gray!10} 
   centroid.$\_$d2.MEI & 0.802 & 0.781 & 0.739 & 0.00017 \\ \rowcolor{gray!10}
   complete.$\_$.EIHI & 0.832 & 0.759 & 0.738 & 0.00018 \\ \rowcolor{gray!10}
   centroid.$\_$.EIHI & 0.798 & 0.778 & 0.734 & 0.00018 \\ \rowcolor{yellow!20}
   kkmeans.$\_$.EIHI & 0.768 & 0.679 & 0.675 & 0.00819 \\ \rowcolor{gray!10}
   centroid$\_$dd2.MEI & 0.607 & 0.697 & 0.575  & 0.00017 \\ \rowcolor{gray!10}
   single.$\_$.EIHI & 0.555 & 0.669 & 0.527  & 0.00020 \\ \rowcolor{gray!10}
   single.$\_$d2.MEI & 0.518 & 0.657 & 0.499 & 0.00015 \\ \rowcolor{gray!10}
   single.$\_$d.MEI & 0.513 & 0.656 & 0.495 & 0.00014 \\ \rowcolor{gray!10}
   single$\_$dd2.MEI & 0.512 & 0.657 & 0.495 & 0.00018 \\ \rowcolor{yellow!20}
   kkmeans.d2.EIHI & 0.501 & 0.503 & 0.495 & 0.00632 \\ \rowcolor{green!10}
   svc.d2.EIHI & 0.501 & 0.534 & 0.495 & 0.00264 \\ \rowcolor{cyan!10} 
   kkmeans.d2.EIHI & 0.501 & 0.523 & 0.495 & 0.00565 \\ \rowcolor{cyan!10}
  kkmeans.dd2.MEI & 0.501 & 0.495 & 0.495 & 0.00566 \\ \rowcolor{orange!20}
   svc.d2.EIHI & 0.501 & 0.535 & 0.495 & 0.00251 \\ 
  spc.d2.EIHI & 0.501 & 0.625 & 0.495 & 0.01183 \\ \rowcolor{gray!10}
  kmeans.d2.EIHI & 0.501 & 0.542 & 0.495 & 0.00115 \\ \rowcolor{magenta!10}
   kmeans.d2.EIHI & 0.501 & 0.542 & 0.495 & 0.00118 \\ \rowcolor{green!10}
   svc.dd2.MEI & 0.500 & 0.498 & 0.495 & 0.00228 \\ \rowcolor{orange!20}
   svc.dd2.MEI & 0.500 & 0.498 & 0.495 & 0.00219 \\\rowcolor{gray!10} 
   kmeans.dd2.MEI & 0.500 & 0.498 & 0.495 & 0.00119 \\ \rowcolor{magenta!10}
   kmeans.dd2.MEI & 0.500 & 0.499 & 0.495 & 0.00125 \\ \rowcolor{gray!10}
   complete.d2.EIHI & 0.500 & 0.598 & 0.495 & 0.00018 \\ \rowcolor{yellow!20}
   kkmeans.dd2.MEI & 0.500 & 0.553 & 0.495 & 0.00699 \\ \rowcolor{gray!10}
   ward.D2.d2.EIHI & 0.500 & 0.560 & 0.495 & 0.00017 \\ \rowcolor{gray!10}
  average.d2.EIHI & 0.500 & 0.623 & 0.495 & 0.00065 \\ \rowcolor{gray!10}
  average.dd2.MEI & 0.500 & 0.586 & 0.495 & 0.00063 \\ \rowcolor{gray!10}
   centroid.d2.EIHI & 0.500 & 0.636 & 0.495 & 0.00013 \\ \rowcolor{gray!10}
   centroid.dd2.MEI & 0.500 & 0.642 & 0.495 & 0.00014 \\ \rowcolor{gray!10}
   complete.dd2.MEI & 0.500 & 0.522 & 0.495 & 0.00018 \\ \rowcolor{gray!10}
   single.d2.EIHI & 0.500 & 0.643 & 0.495 & 0.00013 \\ \rowcolor{gray!10}
   single.dd2.MEI & 0.500 & 0.651 & 0.495 & 0.00019 \\ 
   spc.dd2.MEI & 0.500 & 0.643 & 0.495 & 0.02696 \\ \rowcolor{gray!10} 
   ward.D2.dd2.MEI & 0.500 & 0.514 & 0.495 & 0.00012 \\ 
   \hline
\end{tabular}
}}\hspace*{-1cm}
\caption{Mean results for \textit{S 1-4} considering Euclidean distance (gray), Mahalanobis distance (pink), a gaussian kernel (yellow), a polynomial kernel (blue), kernel k-means for initialization (green) and k-means for initialization (orange).}
\label{tables3}
\end{table}

These results are compared to those obtained from applying functional k-means and test based k-means techniques, which are showed in Tables \ref{tabm3} and \ref{tablez3}. In functional k-means each row represents a different distance between generalized Mahalanobis distance ($d\rho$), truncated Mahalanobis distance ($dk$) and Euclidean distance ($L^2$). In test based k-means, each row stands for a different initialization. When considering the first procedure, $L^2$ distance provides the best RI, 0.847, that is close to those methods with a small value of $\rho$. Nevertheless, when considering $\rho$ equal to 0.02 the execution time is the double of that for $L^2$ distance. 

\begin{table}[ht]
\centering
\footnotesize
 \begin{tabular}{lcccc}
  \hline
 & Purity & Fmeasure & RI & Time \\ 
  \hline
  $L^2$ & 0.917 & 0.846 & 0.847 & 0.72276  \\
  $d\rho$, $\rho=0.02$ & 0.916 & 0.845 & 0.846 & 1.69700  \\
  $d\rho$, $\rho=1$ & 0.916 & 0.844 & 0.846 & 1.73552  \\
  $d\rho$, $\rho=0.001$ & 0.914 & 0.842 & 0.843 & 1.73665 \\ 
  $d\rho$, $\rho=100$ & 0.832 & 0.777 & 0.773 & 2.15567 \\ 
  $d\rho$, $\rho=1e+08$ & 0.863 & 0.772 & 0.772 & 1.87334 \\
  $dk$, $k=2$  & 0.791 & 0.713 & 0.713 & 0.84618  \\
  $dk$, $k=3$ & 0.724 & 0.646 & 0.643 & 0.71322 \\
   \hline 
\end{tabular}
 \caption{Mean values of Purity, F-measure, Rand Index and execution time for the functional $k$-means procedure (\citet{martino}) with truncated Mahalanobis distance, generalized Mahalanobis distance and $L^2$ distance to simulated data from S 1-4.}
    \label{tabm3}
\end{table}

\begin{table}[ht]
\centering
\footnotesize
\begin{tabular}{lcccc}
  \hline
 & Purity & Fmeasure & RI & Time \\ 
  \hline 
  kmeans $++$ & 0.500 & 0.507 & 0.495 & 0.32627  \\
  hclust & 0.500 & 0.498 & 0.495 & 0.20851  \\ 
   kmeans & 0.500 & 0.498 & 0.495 & 0.20502  \\ 
   random & 0.500 & 0.502 & 0.495 & 0.27125 \\ 
   \hline
\end{tabular}
 \caption{Mean values of Purity, F-measure, Rand Index and execution time for the test based $k$-means procedure (\citet{zambom2019}) with four different initialization to simulated data from S 1-4.}
\label{tablez3}
\end{table}

When applying test based k-means the method is not able to distinguish between the two groups, since all the considered metrics get values close to 0.5 in all cases.

Our methodology, besides being the one obtaining the best values in terms of metrics, is the fastest. In this case more than 300 times faster than the best method chosen with functional k-means. The difference in the execution times is key to say that the proposed methodology is a very good alternative to the existing ones to cluster functional data.

Results from the other seven scenarios, whose tables appear in the Supplementary Material, are competitive in terms of metrics and always obtain better execution times.




On the other hand, in order to obtain a further simulation study, data obtained as explained in \citet{martino} is considered, because this simulation type is specially created for testing clustering techniques for functional data.

This type of simulated data consist of taking two functional samples defined in $[0,1]$, having continuous trajectories and which are generated by independent stochastic process in $L^2(I)$. In this case, each curve has $150$ equidistant observations in the interval $[0,1]$. We generate $100$ functions, $50$ from Model $10$ and $50$ from Model $i$, $i=11,12$, obtaining two different functional samples. These two scenarios will be referred as \textit{S 10-11} and \textit{S 10-12} respectively.

The three different models defined for these simulations are explained below.

\begin{itemize}
\item[]\textbf{Model 10.} The first 50 functions are generated as follows: $$X_{10}(t)=E_2(t)+ \sum_{k=1}^{100} Z_k\sqrt{\rho_k}\theta_k(t),$$ where $E_2(t)=t(1-t)$ is the mean function, $\{Z_k, k=1,...,100\}$ is a standard normal variable, $\{ \rho_k,k\geq 1 \}$ is a positive real numbers sequence defined as $$\rho_k = \left\{
	       \begin{array}{lll}
		 \frac{1}{k+1}      & if & k \in \{1,2,3\}, \\
		 \frac{1}{{(k+1)}^2} & if & k \geq 4,
	       \end{array}
	     \right. $$
in such a way that the values of $\rho_k$ are chosen to decrease quicker when $k\geq 4$ in order to have most of the variance explained by the first three principal components. The sequence $\{\theta_k, k\geq 1\}$ is an orthonormal basis of $L^2(I)$ defined as$$\theta_k(t) = \left\{
	       \begin{array}{lllll}
		 \mathbbm{1}_{[0,1]}(t)     & if & k=1, &  \\
		 \sqrt{2}\sin{(k\pi t)}\mathbbm{1}_{[0,1]}(t) & if & k \geq 2, & k \ even,\\
		 \sqrt{2}\cos{((k-1)\pi t)}\mathbbm{1}_{[0,1]}(t) & if & k \geq 3, & k \ odd.
	       \end{array}
	     \right. $$
	    
The next two models are defined in the same way but changing in each case the term which is added to $E_2(t)$ in Model 10. Moreover, the standard normal variables generated for the following two models differs from those of the last model.  
\item[]\textbf{Model 11.} $X_{11}(t)=E_3(t)+ \displaystyle \sum_{k=1}^{100} Z_k\sqrt{\rho_k}\theta_k(t),$ where $E_3(t)=E_2(t)+\displaystyle \sum_{k=1}^3\sqrt{\rho_k}\theta_k(t)$  
\item[]\textbf{Model 12.} $X_{12}(t)=E_4(t)+ \displaystyle \sum_{k=1}^{100} Z_k\sqrt{\rho_k}\theta_k(t),$ where $E_4(t)=E_2(t)+\displaystyle \sum_{k=4}^{100}\sqrt{\rho_k}\theta_k(t)$  
\end{itemize}

As before, we consider the data smoothed with a cubic B-spline basis to remove noise and to be able to use first and second derivatives of the data.

\begin{figure}[ht]
    \centering
    \includegraphics[scale=0.55]{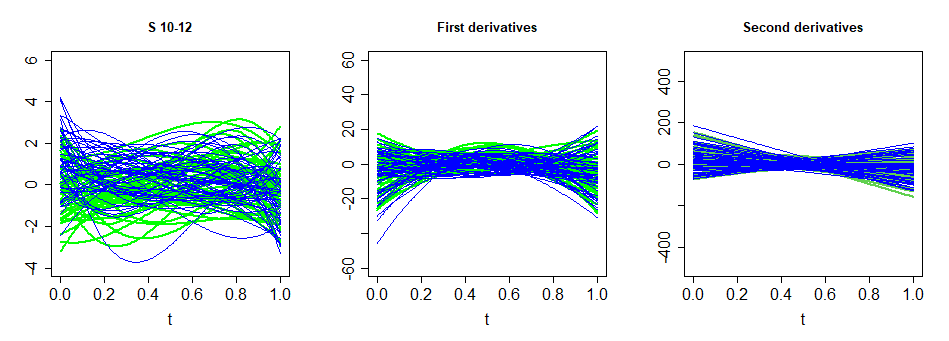}
    \caption{A sample generated from \textit{S 10-12}. Original data, first and second derivatives curves.}
    \label{figsc10f}
\end{figure}

\begin{figure}[ht]
    \centering
    \includegraphics[scale=0.55]{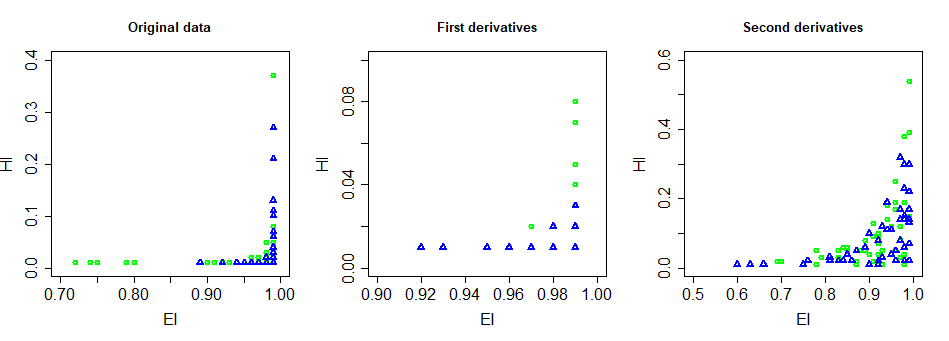}
     \caption{Scatter plots of the epigraph index (EI) and the hypograph index (HI) of the original data simulated from Model 10 and 12 (left panel), first derivatives (second panel) and second derivatives (right panel).}
    \label{figsc10ind}
\end{figure}

\begin{figure}[ht]
    \centering
    \includegraphics[scale=0.55]{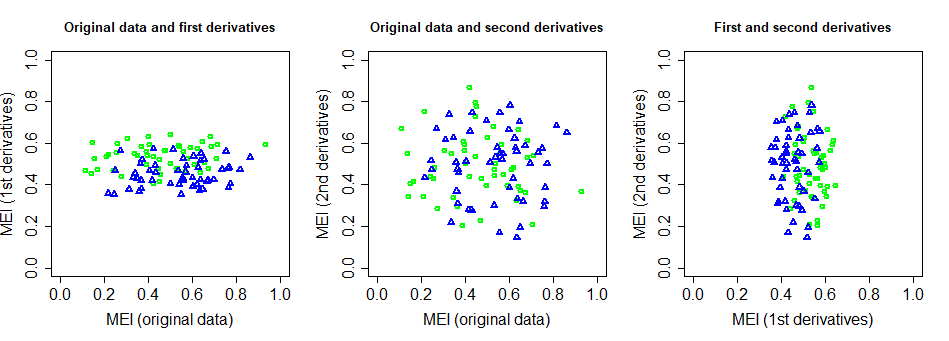}
    \caption{A sample generated from \textit{S 10-12}. Scatter plots of different combinations of MEI.  Original data and first derivatives (left panel), original data and second derivatives (center panel) and first and second derivatives (right panel).}
    \label{figsc10Mind}
\end{figure}

When data simulated from \textit{S 10-12} is smoothed, the curves cross a lot between themselves, and also their derivatives (Figure \ref{figsc10f}). When applying the epigraph and the hypograph indexes to these sets of curves, again the difference between groups is negligible (Figure \ref{figsc10ind}). Nevertheless, looking at Figure \ref{figsc10Mind}, when applying the MEI the difference between groups is now much more clear. The best result in Table \ref{tables10} is achieved by applying kernel k-means with a polynomial kernel on the generalized epigraph index of the first and second derivatives, obtaining a RI of 0.919. Moreover, when applying the same technique with the same set of data (first and second derivatives) but now adding the epigraph and hypograph indexes, the same RI is obtained. This combination is not shown since six different variables are involved in it.

\begin{table}[!htbp]
\vspace{-1cm}
\hspace*{-2cm}
\centering
\parbox{10cm}{
\scalebox{0.9}{
\footnotesize \begin{tabular}{lcccc}
\hline 
 & Purity & Fmeasure & RI & Time \\ 
  \hline \rowcolor{cyan!10}
kkmeans.dd2.MEI & 0.957 & 0.918 & 0.919 & 0.00423 \\ \rowcolor{cyan!10}
   kkmeans.dd2.EIHIMEI & 0.957 & 0.918 & 0.918 & 0.00466 \\ \rowcolor{orange!12}
   svc.dd2.EIHIMEI & 0.956 & 0.916 & 0.917 & 0.00276 \\ \rowcolor{green!10}
  svc.dd2.MEI & 0.956 & 0.916 & 0.917 & 0.00185 \\ \rowcolor{gray!10}
  kmeans.dd2.EIHIMEI & 0.956 & 0.916 & 0.917 & 0.00100 \\ \rowcolor{gray!10}
  kmeans.dd2.MEI & 0.956 & 0.916 & 0.917 & 0.00082 \\ \rowcolor{green!10}
 svc.dd2.EIHIMEI & 0.956 & 0.916 & 0.914 & 0.00273 \\ \rowcolor{gray!10}
 ward.D2.dd2.EIHIMEI & 0.939 & 0.889 & 0.888 & 0.00015 \\ \rowcolor{gray!10}
 ward.D2.dd2.MEI & 0.939 & 0.889 & 0.888 & 0.00009 \\ \rowcolor{gray!10}
  average.dd2.MEI & 0.921 & 0.882 & 0.874 & 0.00009 \\ 
 spc.dd2.MEI & 0.909 & 0.885 & 0.869 & 0.02349 \\\rowcolor{yellow!20}
  kkmeans.dd2.EIHIMEI & 0.913 & 0.855 & 0.855 & 0.00808 \\ \rowcolor{green!10}
  svc.d2.EIHIMEI & 0.915 & 0.843 & 0.844 & 0.00226 \\ \rowcolor{yellow!20}
   kkmeans.dd2.MEI & 0.890 & 0.825 & 0.824 & 0.00711 \\ \rowcolor{cyan!10}
  kkmeans.d.EIHIMEI & 0.889 & 0.803 & 0.804 & 0.00398 \\ \rowcolor{gray!10}
   average.d2.EIHIMEI & 0.844 & 0.789 & 0.767 & 0.00011 \\ \rowcolor{gray!10}
  centroid.dd2.EIHIMEI & 0.735 & 0.784 & 0.705 & 0.00014 \\ \rowcolor{gray!10}
   centroid.dd2.MEI & 0.735 & 0.784 & 0.705 & 0.00016 \\ \rowcolor{yellow!20}
  kkmeans.d.EIHIMEI & 0.699 & 0.608 & 0.604 & 0.00684 \\ \rowcolor{cyan!10}
  kkmeans.d2.EIHI & 0.540 & 0.497 & 0.500 & 0.00226 \\ \rowcolor{gray!10}
  complete.$\_$d.MEI & 0.538 & 0.530 & 0.499 & 0.00012 \\ \rowcolor{gray!10}
  ward.D2$\_$dd2.MEI & 0.536 & 0.521 & 0.499 & 0.00015 \\ \rowcolor{gray!10}
  ward.D2.$\_$d.MEI & 0.536 & 0.514 & 0.498 & 0.00010 \\ \rowcolor{cyan!10}
  kkmeans.$\_$d2.MEI & 0.535 & 0.496 & 0.498 & 0.00402 \\ \rowcolor{gray!10}
  complete$\_$dd2.MEI & 0.536 & 0.525 & 0.498 & 0.00050 \\ \rowcolor{cyan!10}
  kkmeans$\_$dd2.MEI & 0.534 & 0.496 & 0.498 & 0.00440 \\ \rowcolor{green!10}
  svc$\_$dd2.MEI & 0.534 & 0.497 & 0.498 & 0.00209 \\ \rowcolor{green!10}
  svc.$\_$dd2.EIHIMEI & 0.533 & 0.498 & 0.498 & 0.00372 \\ \rowcolor{orange!20}
  svc.$\_$d2.MEI & 0.534 & 0.497 & 0.498 & 0.00192 \\ \rowcolor{orange!20}
 svc$\_$dd2.MEI & 0.533 & 0.497 & 0.498 & 0.00194 \\\rowcolor{green!10} 
  svc.$\_$d.MEI & 0.533 & 0.497 & 0.498 & 0.00201 \\ \rowcolor{gray!10}
  kmeans$\_$dd2.MEI & 0.533 & 0.497 & 0.498 & 0.00096 \\ \rowcolor{magenta!10}
  kmeans$\_$dd2.MEI & 0.533 & 0.497 & 0.498 & 0.00091 \\ 
  \rowcolor{magenta!10} 
kmeans.$\_$d.MEI & 0.533 & 0.497 & 0.498 & 0.00088 \\   \end{tabular}
   }}
\parbox{9cm}{
\footnotesize
\scalebox{0.9}{
\begin{tabular}[t]{lcccc}
  \hline
 & Purity & Fmeasure & RI & Time \\ 
  \hline \rowcolor{gray!10} 
kmeans.$\_$d2.MEI & 0.533 & 0.497 & 0.498 & 0.00084 \\ \rowcolor{magenta!10}
  kmeans.$\_$d2.MEI & 0.533 & 0.497 & 0.498 & 0.00091 \\ \rowcolor{green!10} 
   svc.$\_$d2.EIHIMEI & 0.532 & 0.498 & 0.498 & 0.00315 \\ \rowcolor{gray!10}
complete.$\_$d2.MEI & 0.534 & 0.526 & 0.498 & 0.00012 \\
 spc.$\_$d.EIHIMEI & 0.530 & 0.577 & 0.497 & 0.01950 \\ \rowcolor{gray!10}
  average.$\_$d.MEI & 0.527 & 0.555 & 0.497 & 0.00049 \\                 
  spc.$\_$d2.MEI & 0.528 & 0.589 & 0.497 & 0.02303 \\
  spc$\_$dd2.MEI & 0.525 & 0.588 & 0.497 & 0.02132 \\ \rowcolor{cyan!10}
  kkmeans.$\_$.EIHI & 0.524 & 0.620 & 0.496 & 0.00629 \\  
  spc.$\_$.EIHI & 0.523 & 0.625 & 0.496 & 0.00607 \\\rowcolor{orange!20} 
  svc.$\_$.EIHI & 0.523 & 0.614 & 0.496 & 0.00208 \\ \rowcolor{green!10}
 svc.$\_$dd2.EIHI & 0.524 & 0.620 & 0.496 & 0.00331 \\ 
  spc.$\_$d.EIHI & 0.523 & 0.622 & 0.496 & 0.00635 \\
  spc.$\_$dd2.EIHI & 0.521 & 0.626 & 0.496 & 0.00591 \\ \rowcolor{gray!10}
 centroid$\_$dd2.MEI & 0.520 & 0.605 & 0.496 & 0.00049 \\ \rowcolor{magenta!10}
kmeans.$\_$.EIHI & 0.519 & 0.639 & 0.496 & 0.00088 \\ \rowcolor{gray!10}
kmeans.$\_$dd2.EIHI & 0.519 & 0.639 & 0.496 & 0.00097 \\ \rowcolor{gray!10}
 ward.D2.$\_$dd2.EIHI & 0.518 & 0.639 & 0.496 & 0.00013 \\ \rowcolor{gray!10}
  average.$\_$dd2.EIHIMEI & 0.51670 & 0.62758 & 0.496 & 0.00016 \\ \rowcolor{gray!10}
complete.$\_$d2.EIHI & 0.514 & 0.647 & 0.495 & 0.00012 \\ \rowcolor{gray!10}
  complete.$\_$dd2.EIHI & 0.514 & 0.647 & 0.495 & 0.00012 \\ \rowcolor{gray!10}
 single.$\_$d.EIHIMEI & 0.511 & 0.653 & 0.495 & 0.00017 \\ \rowcolor{gray!10} 
centroid.$\_$.EIHI & 0.511 & 0.653 & 0.495 & 0.00011 \\ \rowcolor{gray!10}
 centroid.$\_$d.EIHI & 0.511 & 0.653 & 0.495 & 0.00014 \\ \rowcolor{gray!10}
single.$\_$dd2.EIHIMEI & 0.511 & 0.654 & 0.495 & 0.00019 \\ \rowcolor{gray!10}
  single.$\_$.EIHI & 0.511 & 0.655 & 0.495 & 0.00055 \\ \rowcolor{gray!10}
 single.$\_$dd2.EIHI & 0.511 & 0.655 & 0.495 & 0.00013 \\ \rowcolor{gray!10}
 single.dd2.MEI & 0.510 & 0.657 & 0.495 & 0.00012 \\ \rowcolor{gray!10}
average.d.EIHI & 0.510 & 0.657 & 0.495 & 0.00009 \\
  \rowcolor{gray!10} 
 centroid.d2.EIHI & 0.510 & 0.657 & 0.495 & 0.00009 \\ \rowcolor{gray!10} 
   complete.d.EIHI & 0.510 & 0.657 & 0.495 & 0.00010 \\ \rowcolor{gray!10}
complete.d2.EIHI & 0.510 & 0.657 & 0.495 & 0.00012 \\ \rowcolor{gray!10}
  single.d2.EIHI & 0.510 & 0.657 & 0.495 & 0.00010 \\ \rowcolor{gray!10}
 single.dd2.EIHI & 0.510 & 0.657 & 0.495 & 0.00010 \\ 
   \hline
\end{tabular}
}}\hspace*{-1cm}
\caption{Mean results for S 10-12 considering Euclidean distance (gray), Mahalanobis distance (pink), a gaussian kernel (yellow), a polynomial kernel (blue), kernel k-means for initialization (green) and k-means for initialization (orange).}
\label{tables10}
\end{table}

These results are compared to those obtained by applying functional k-means procedure (Table \ref{tabm10}). In this case, the best distance is the Mahalanobis distance with a big value of $\rho$, $\rho=1e+08$, obtaining a RI of 0.718, which is small compared to the value of 0.919 obtained with our strategy. Besides, our methodology spends 0.00423 seconds per iteration, while their procedure spends 7.9055 seconds.

\begin{table}[ht]
\centering
\footnotesize
 \begin{tabular}{lcccc}
  \hline
 & Purity & Fmeasure & RI & Time \\ 
  \hline
  $d\rho$, $\rho=1e+08$ & 0.831 & 0.718 & 0.718 & 7.9055 \\ 
  $d\rho$, $\rho=100$ & 0.637 & 0.554 & 0.548 & 10.1940 \\ 
  $d\rho$, $\rho=0.02$ & 0.551 & 0.504 & 0.502 & 9.2982\\ 
  $d\rho$, $\rho=1$ & 0.549 & 0.503 & 0.502 & 9.8511\\ 
  $d\rho$, $\rho=0.001$ & 0.548 & 0.503 & 0.502 & 9.9013\\ 
  $L^2$ & 0.547 & 0.502 & 0.502 & 1.3529\\ 
  $dk$, $k=3$ & 0.543 & 0.503 & 0.501 & 1.1880\\
  $dk$, $k=2$ & 0.541 & 0.501 & 0.500 & 0.53262\\
   \hline 
\end{tabular}
 \caption{Mean values of Purity, F-measure, Rand Index and execution time for the functional $k$-means procedure (\citet{martino}) with truncated Mahalanobis distance, generalized Mahalanobis distance and $L^2$ distance to simulated data from S 10-12.}
    \label{tabm10}
\end{table}

When applying test based k-means to this type of simulated data, results in Table \ref{tablez10} show that any of the initialization strategies are able to distinguish between groups, obtaining close values to 0.5 for all metrics.

\begin{table}[ht]
\centering
\footnotesize
\begin{tabular}{lcccc}
  \hline
 & Purity & Fmeasure & RI & Time \\ 
  \hline 
  kmeans & 0.545 & 0.502 & 0.501 & 0.30259  \\
   hclust & 0.542 & 0.502 & 0.500 & 0.39506  \\ 
   random & 0.539 & 0.501 & 0.499 & 0.53736 \\ 
   kmeans $++$ & 0.536 & 0.503 & 0.499 & 0.99704  \\ 
   \hline
\end{tabular}
 \caption{Mean values of Purity, F-measure, Rand Index and execution time for the test based $k$-means procedure (\citet{zambom2019}) with four different initialization to simulated data from S 10-12.}
\label{tablez10}
\end{table}

We conclude that the new methodology obtains the best results in terms of metrics and execution time.


\subsection{Simulation Study B: Three clusters}

In this case, we consider three different scenarios coming from three different groups. This simulation study previously appeared in \citet{zambom2019}. Each data set is composed by 150 curves, belonging 50 of them to each of the three clusters. Each curve has 100 equidistant points defined in the interval $[0, \frac{\pi}{3}]$.

Each scenario is composed by 50 functions from three different models, each of them of the form:

$$X(t) = Y(t)+\epsilon$$

The nine different models are defined as follows:
\begin{itemize}
    \item[] \textbf{Model 13.} $X_{13}(t) = \frac{1}{1.3}\sin(1.3 t)+t^3+a+0.3+\epsilon_1$
    \item[] \textbf{Model 14.} $X_{14}(t) = \frac{1}{1.2}\sin(1.3 t)+t^3+a+1+\epsilon_1$
    \item[] \textbf{Model 15.} $X_{15}(t) = \frac{1}{4}\sin(1.3 t)+t^3+a+0.2+\epsilon_1$
    \item[] \textbf{Model 16.} $X_{16}(t) = \sin(1.5 \pi t)+\cos(\pi t^2)+b+1.1+\epsilon_1$
    \item[] \textbf{Model 17.} $X_{17}(t) = \sin(1.7 \pi t)+\cos(\pi t^2)+b+1.5+\epsilon_1$
    \item[] \textbf{Model 18.} $X_{18}(t) = \sin(1.9 \pi t)+\cos(\pi t^2)+b+2.2+\epsilon_1$
    \item[] \textbf{Model 19.} $X_{19}(t) = \frac{1}{1.8}\exp(1.1 t)-t^3+a+\epsilon_2$
    \item[] \textbf{Model 20.} $X_{20}(t) = \frac{1}{1.7}\exp(1.4 t)-t^3+a+\epsilon_2$
    \item[] \textbf{Model 21.} $X_{21}(t) = \frac{1}{1.5}\exp(1.5 t)-t^3+a+\epsilon_2$
\end{itemize}

where $a\sim U(\frac{-1}{4}, \frac{1}{4})$, $b\sim U(\frac{-1}{2}, \frac{1}{2})$ and $\epsilon_1\sim N(2,0.4^2)$, $\epsilon_2\sim N(2,0.4^2)$ for each curve.

This way, \textit{S 13-14-15} is composed by 50 functions from Model 13, 50 functions from Model 14 and 50 of them from Model 15, \textit{S 16-17-18} is composed by 50 functions of each Model 16,17 and 18, and \textit{S 19-20-21} is created in the same way from models 19, 20 and 21.

\begin{figure}[ht]
    \centering
    \includegraphics[scale=0.55]{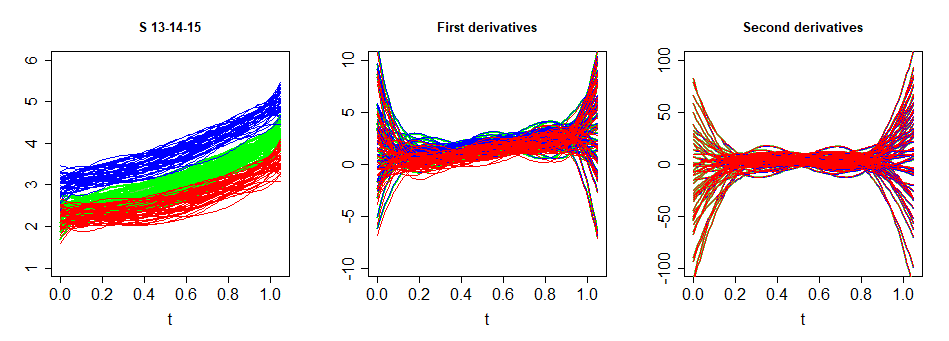}
    \caption{A sample generated from S 13-14-15. Original data (left panel), first (center panel) and second derivatives curves (right panel).}
    \label{figsc11f}
\end{figure}

\begin{figure}[ht]
    \centering
    \includegraphics[scale=0.55]{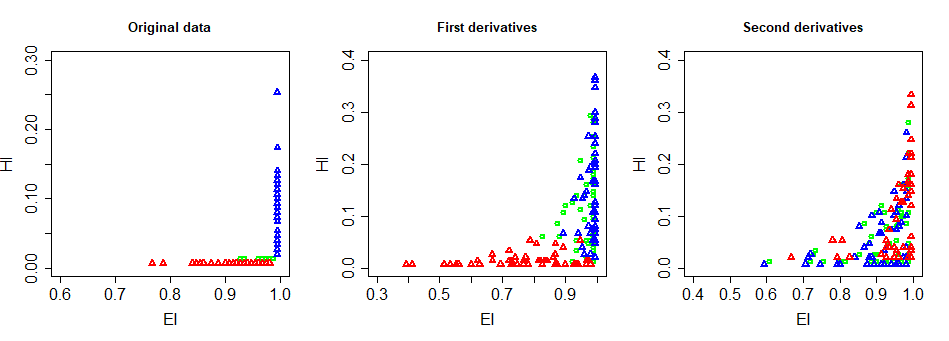}
     \caption{Scatter plots of the epigraph index (EI) and the hypograph index (HI) of the original data simulated from Model 13, 14 and 15 (left panel), first derivatives (center panel) and second derivatives (right panel).}
    \label{figsc11ind}
\end{figure}

\begin{figure}[ht]
    \centering
    \includegraphics[scale=0.55]{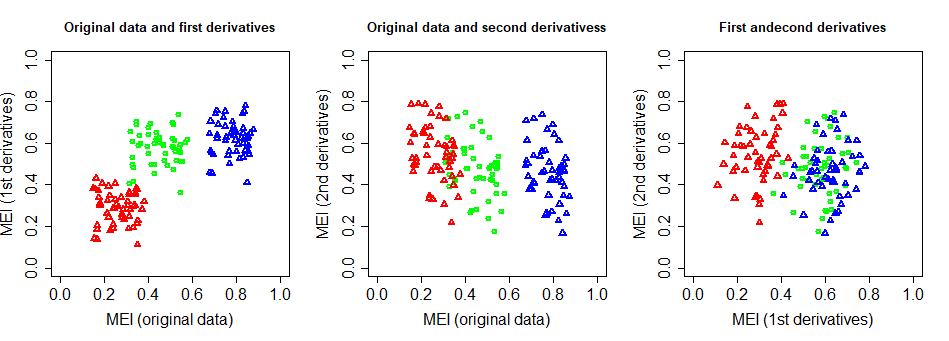}
    \caption{A sample generated from S 13-14-15. Scatter plots of different combinations of MEI. Original data and first derivatives (left panel), original data and second derivatives (center panel) and first and second derivatives (right panel).}
    \label{figsc11Mind}
\end{figure}

Data considered for \textit{S 13-14-15} is shown in Figure \ref{figsc11f}. We observe that functions in green and red intertwine a lot. Moreover, we observe that when applying the epigraph and the hypograph indexes (Figure \ref{figsc11ind}) there is a clear difference between two groups, but the green one is overlapped with the other two. Nevertheless, when considering the modified epigraph index (Figure \ref{figsc11Mind}) it seems that the difference between the three groups are much more evident.

When applying our methodology using k-means with both Euclidean or Mahalanobis distances to the first and second derivatives of the generalized epigraph index ends in the best result in Table \ref{tables11}. (RI=0.983 and ET=0.003 seconds for both distances). When considering functional k-means, the best method in Table \ref{tabm11} is the one with a small value of $\rho$, $\rho=0.001$, (RI=0.928 and ET=6.50802 seconds). And when applying test based k-means, the best result in Table \ref{tablez11} is obtained when initializing the process with k-means$++$. (RI=0.944 and ET=0.99653).  
\begin{table}[htbp]
\vspace{-1cm}
\hspace*{-2cm}
\centering
\parbox{10cm}{

\footnotesize 
\scalebox{0.9}{
\begin{tabular}{lcccc}
\hline 
 & Purity & Fmeasure & RI & Time \\ 
  \hline  \rowcolor{gray!10}
kmeans$\_$dd2.MEI & 0.987 & 0.974 & 0.983 & 0.00257 \\ \rowcolor{magenta!10}
 kmeans$\_$dd2.MEI & 0.987 & 0.974 & 0.983 & 0.00267 \\ \rowcolor{green!10}
  svc.$\_$d.MEI & 0.986 & 0.974 & 0.982 & 0.00694 \\ \rowcolor{gray!10} 
  kmeans.$\_$d.MEI & 0.986 & 0.973 & 0.982 & 0.00216 \\ \rowcolor{magenta!10}
  kmeans.$\_$d.MEI & 0.986 & 0.973 & 0.982 & 0.00222 \\ \rowcolor{orange!20}
 svc.$\_$d.MEI & 0.986 & 0.973 & 0.982 & 0.00590 \\ \rowcolor{orange!20} 
  svc$\_$dd2.MEI & 0.983 & 0.970 & 0.980 & 0.00667 \\ \rowcolor{green!10}
  svc$\_$dd2.MEI & 0.980 & 0.966 & 0.977 & 0.00680 \\ \rowcolor{gray!10}
   ward.D2.$\_$d.MEI & 0.976 & 0.954 & 0.969 & 0.00047 \\ \rowcolor{gray!10}
  average.$\_$d.MEI & 0.974 & 0.954 & 0.969 & 0.00195 \\ \rowcolor{gray!10}
   ward.D2$\_$dd2.MEI & 0.974 & 0.950 & 0.967 & 0.00039 \\ \rowcolor{gray!10}
   centroid.$\_$d.MEI & 0.969 & 0.952 & 0.966 & 0.00037 \\ \rowcolor{gray!10}
  average$\_$dd2.MEI & 0.961 & 0.940 & 0.958 & 0.00038 \\ \rowcolor{gray!10}
   complete.$\_$d.MEI & 0.962 & 0.936 & 0.957 & 0.00195 \\ \rowcolor{yellow!20}
   kkmeans$\_$dd2.MEI & 0.958 & 0.934 & 0.955 & 0.02536 \\ \rowcolor{cyan!10}
   kkmeans.$\_$d.MEI & 0.953 & 0.917 & 0.945 & 0.01830 \\ 
  spc.$\_$d.MEI & 0.922 & 0.919 & 0.936 & 0.09266 \\ 
  spc$\_$dd2.MEI & 0.919 & 0.910 & 0.930 & 0.08338 \\ \rowcolor{yellow!20}
  kkmeans.$\_$d.MEI & 0.930 & 0.897 & 0.928 & 0.02682 \\ \rowcolor{cyan!10}
   kkmeans$\_$dd2.MEI & 0.929 & 0.884 & 0.922 & 0.02367 \\ \rowcolor{gray!10}
   complete$\_$dd2.MEI & 0.860 & 0.825 & 0.877 & 0.00037 \\ \rowcolor{gray!10}
   kmeans.$\_$d2.MEI & 0.845 & 0.764 & 0.844 & 0.00252 \\ \rowcolor{magenta!10}
   kmeans.$\_$d2.MEI & 0.845 & 0.764 & 0.844 & 0.00266 \\ \rowcolor{gray!10}
   ward.D2.$\_$d2.MEI & 0.828 & 0.758 & 0.837 & 0.00038 \\ \rowcolor{green!10}
   svc.$\_$d2.MEI & 0.826 & 0.749 & 0.830 & 0.00798 \\ \rowcolor{orange!20}
   svc.$\_$d2.MEI & 0.812 & 0.736 & 0.820 & 0.00760 \\ \rowcolor{yellow!20}
   kkmeans.$\_$d2.MEI & 0.804 & 0.723 & 0.814 & 0.03077 \\ \rowcolor{cyan!10}
   kkmeans.$\_$d2.MEI & 0.785 & 0.718 & 0.806 & 0.02304 \\ 
   spc.$\_$d2.MEI & 0.737 & 0.742 & 0.792 & 0.08898 \\ \rowcolor{gray!10}
   average.$\_$d2.MEI & 0.714 & 0.731 & 0.787 & 0.00034 \\ \rowcolor{gray!10}
   complete.$\_$d2.MEI & 0.732 & 0.677 & 0.770 & 0.00112 \\ \rowcolor{gray!10}
  centroid.$\_$d2.MEI & 0.669 & 0.730 & 0.762 & 0.00040 \\ \rowcolor{gray!10}
  centroid$\_$dd2.MEI & 0.680 & 0.736 & 0.755 & 0.00042 \\ \rowcolor{gray!10}
   kmeans.dd2.MEI & 0.644 & 0.604 & 0.73578 & 0.00276 \\ \rowcolor{magenta!10}
  kmeans.dd2.MEI & 0.644 & 0.604 & 0.735 & 0.00297 \\ \rowcolor{orange!20}
   svc.dd2.MEI & 0.641 & 0.601 & 0.731 & 0.00877 \\ 
       \end{tabular}}
   }
\parbox{9cm}{
\footnotesize
\scalebox{0.9}{
\begin{tabular}[t]{lcccc}
  \hline
 & Purity & Fmeasure & RI & Time \\ 
  \hline \rowcolor{green!10}
  svc.dd2.MEI & 0.640 & 0.600 & 0.727 & 0.00858 \\ \rowcolor{gray!10}
  ward.D2.dd2.MEI & 0.631 & 0.592 & 0.717 & 0.00044 \\ \rowcolor{cyan!10}
  kkmeans.dd2.MEI & 0.631 & 0.574 & 0.708 & 0.02201 \\ \rowcolor{gray!10}
   single.$\_$d.MEI & 0.641 & 0.715 & 0.704 & 0.00122 \\  \rowcolor{gray!10}
average.dd2.MEI & 0.626 & 0.636 & 0.703 & 0.00034 \\ \rowcolor{yellow!20}
  kkmeans.dd2.MEI & 0.606 & 0.542 & 0.690 & 0.03032 \\ \rowcolor{gray!10}
   complete.dd2.MEI & 0.606 & 0.562 & 0.684 & 0.00038 \\ 
   spc.dd2.MEI & 0.582 & 0.601 & 0.650 & 0.09179 \\ \rowcolor{cyan!10}
  kkmeans.d.EIHI & 0.570 & 0.482 & 0.634 & 0.02055 \\ \rowcolor{gray!10}
  ward.D2.d.EIHI & 0.562 & 0.491 & 0.622 & 0.00041 \\ \rowcolor{orange!20}
  svc.d.EIHI & 0.564 & 0.476 & 0.621 & 0.01003 \\ \rowcolor{gray!10}
  kmeans.d.EIHI & 0.561 & 0.472 & 0.621 & 0.00251 \\ \rowcolor{magenta!10}
  kmeans.d.EIHI & 0.561 & 0.472 & 0.621 & 0.00252 \\ \rowcolor{green!10}
  svc.d.EIHI & 0.561 & 0.472 & 0.620 & 0.00998 \\ \rowcolor{yellow!20}
   kkmeans.d.EIHI & 0.521 & 0.437 & 0.613 & 0.02567 \\ \rowcolor{gray!10}
   single.$\_$d2.MEI & 0.531 & 0.628 & 0.580 & 0.00038 \\ 
   spc.d.EIHI & 0.549 & 0.528 & 0.579 & 0.09115 \\ \rowcolor{gray!10} 
   centroid.dd2.MEI & 0.519 & 0.589 & 0.566 & 0.00041 \\ \rowcolor{gray!10}
  complete.d.EIHI & 0.528 & 0.488 & 0.558 & 0.00044 \\ \rowcolor{yellow!20}
   kkmeans.d2.EIHI & 0.396 & 0.343 & 0.553 & 0.03102 \\ \rowcolor{gray!10}
   average.d.EIHI & 0.530 & 0.528 & 0.547 & 0.00033 \\ \rowcolor{cyan!10}
   kkmeans.d2.EIHI & 0.411 & 0.368 & 0.540 & 0.02190 \\ \rowcolor{gray!10}
  kmeans.d2.EIHI & 0.404 & 0.384 & 0.522 & 0.00259 \\ \rowcolor{magenta!10}
  kmeans.d2.EIHI & 0.404 & 0.384 & 0.522 & 0.00289 \\ \rowcolor{green!10}
   svc.d2.EIHI & 0.404 & 0.384 & 0.522 & 0.01000 \\ \rowcolor{orange!20}
   svc.d2.EIHI & 0.404 & 0.384 & 0.522 & 0.01163 \\ \rowcolor{gray!10}
   ward.D2.d2.EIHI & 0.400 & 0.386 & 0.518 & 0.00038 \\ \rowcolor{gray!10}
   centroid.d.EIHI & 0.491 & 0.512 & 0.498 & 0.00114 \\ \rowcolor{gray!10}
   single$\_$dd2.MEI & 0.447 & 0.566 & 0.469 & 0.00040 \\ \rowcolor{gray!10}
   complete.d2.EIHI & 0.387 & 0.425 & 0.468 & 0.00046 \\ 
   spc.d2.EIHI & 0.378 & 0.432 & 0.449 & 0.07513 \\ \rowcolor{gray!10}
   average.d2.EIHI & 0.369 & 0.447 & 0.427 & 0.00032 \\ \rowcolor{gray!10}
  centroid.d2.EIHI & 0.365 & 0.456 & 0.412 & 0.00039 \\ \rowcolor{gray!10}
   single.d.EIHI & 0.370 & 0.490 & 0.360 & 0.00126 \\ \rowcolor{gray!10}
  single.d2.EIHI & 0.345 & 0.484 & 0.354 & 0.00040 \\ \rowcolor{gray!10}
   single.dd2.MEI & 0.352 & 0.488 & 0.351 & 0.00044 \\ 
   \hline
\end{tabular}}
}\hspace*{-1cm}
\caption{Mean results for S 13-14-15 considering Euclidean distance (gray), Mahalanobis distance (pink), a gaussian kernel (yellow), a polynomial kernel (blue), kernel k-means for initialization (green) and k-means for initialization (orange).}
\label{tables11}
\end{table}

\begin{table}[ht]
\centering
\footnotesize
 \begin{tabular}{lcccc}
  \hline
 & Purity & Fmeasure & RI & Time \\ 
  \hline
  $d\rho$, $\rho=0.001$ & 0.936 & 0.894 & 0.928 & 6.50802 \\
  $d\rho$, $\rho=0.02$ & 0.934 & 0.890 & 0.925 & 6.28237  \\
  $L^2$ & 0.934 & 0.887 & 0.923 & 1.38996  \\
  $d\rho$, $\rho=1$ & 0.927 & 0.885 & 0.921 & 6.39378 \\
  $d\rho$, $\rho=100$ & 0.682 & 0.662 & 0.754 & 5.67288 \\ 
  $dk$, $k=2$ & 0.659 & 0.606 & 0.735 & 1.54468 \\
  $d\rho$, $\rho=1e+08$ &  0.719 & 0.590 & 0.725 & 7.95828 \\
  $dk$, $k=3$ & 0.605 & 0.548 & 0.695 & 1.59957 \\
   \hline 
\end{tabular}
 \caption{Mean values of Purity, F-measure, Rand Index and execution time for the functional $k$-means procedure (\citet{martino}) with truncated Mahalanobis distance, generalized Mahalanobis distance and $L^2$ distance to simulated data from S 13-14-15}
    \label{tabm11}
\end{table}

\begin{table}[ht]
\centering
\footnotesize
\begin{tabular}{lcccc}
  \hline
 & Purity & Fmeasure & RI & Time \\ 
  \hline 
kmeans $$++$$  & 0.955 & 0.915 & 0.944 & 0.99653\\ 
  kmeans & 0.953 & 0.912 & 0.942 & 4.96029\\ 
  random & 0.952 & 0.910 & 0.940 & 1.05298\\ 
  hclust & 0.947 & 0.903 & 0.936 & 4.98203\\ 
   \hline
\end{tabular}
 \caption{Mean values of Purity, F-measure, Rand Index and execution time for the test based $k$-means procedure (\citet{zambom2019}) with four different initialization to simulated data from S 13-14-15.}
\label{tablez11}
\end{table}

In summary, the three methodologies provide accurate results, but the proposed procedure is the one obtaining the best values in terms of metrics and execution time. 

Results obtained for \textit{S 16-17-18} and \textit{S 19-20-21} are shown 
in  the Supplementary Material.
Regarding data simulated for these two scenarios, our methodology gets the best execution time for both of them, obtaining competitive results in terms of metrics. 

\FloatBarrier
\section{Application to real data}
\label{sec:realdata}
From the simulation study conducted in Section \ref{sec:sim}, we conclude that, hierarchical methods perform worse than the other clustering techniques concerning k-means. Because of that, and to reduce the calculations, in this section these methods are omitted.

\subsection{Case study: Growth data set}
We have applied our methodology to a popular real data set in the FDA literature: the Berkeley growth study. This is a classical data set included in \citet{ramsay} and available in the `fda' R-package. It contains the heights of 93 people from age 1 to 18 (54 girls and 39 boys).

\begin{figure}[ht]
    \centering
    \includegraphics[scale=0.4]{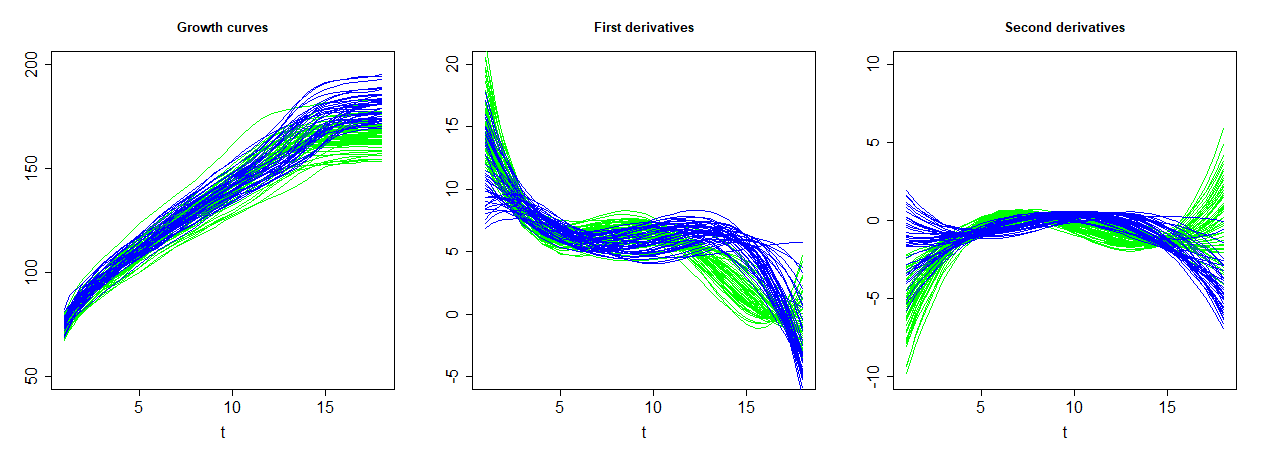}
    \caption{Growth curves (girls in green and boys in blue) for the original data (left panel) the first derivatives (center panel) and the second derivatives (right panel).}
    \label{figscgrf}
\end{figure}

\begin{figure}[ht]
    \centering
    \includegraphics[scale=0.4]{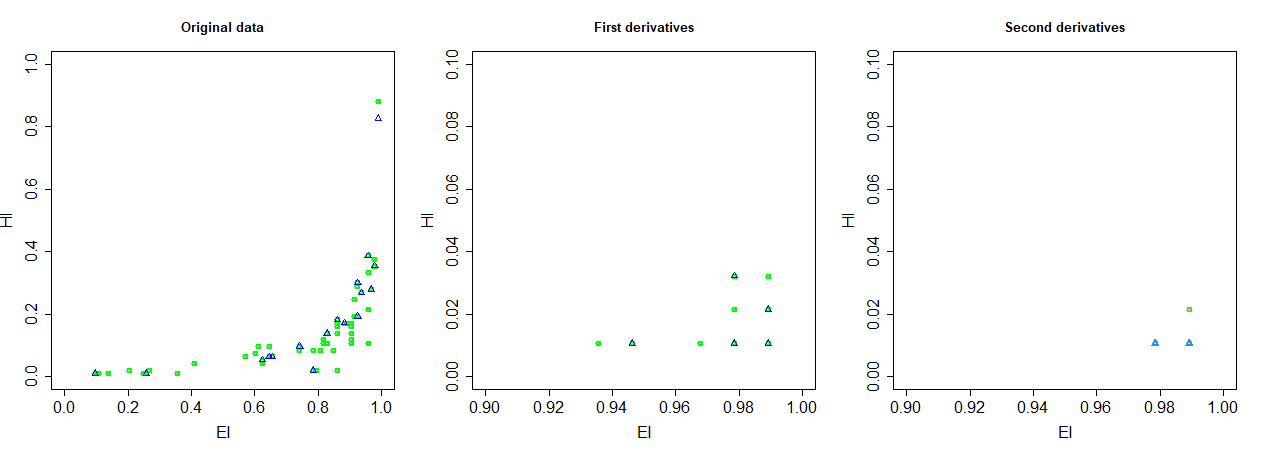}
     \caption{Scatter plots of the epigraph index (EI) and the hypograph index (HI) of the growth curves original data (left panel), first derivatives (second panel) and second derivatives (right panel).}
    \label{figscgrind}
\end{figure}

\begin{figure}[ht]
    \centering
    \includegraphics[scale=0.4]{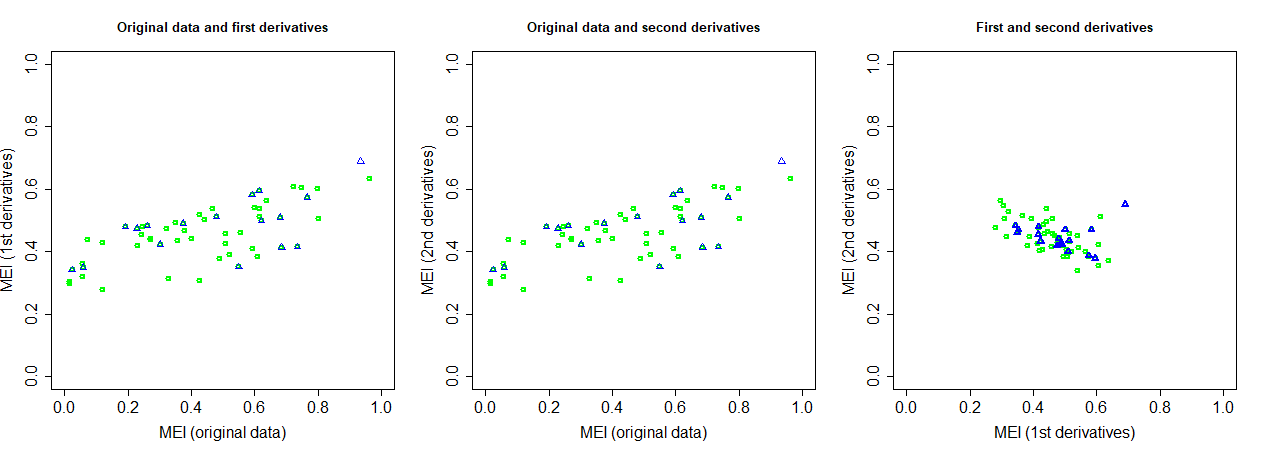}
    \caption{Growth curves. Scatter plots of different combinations of MEI. Original data and first derivatives (left panel), original data and second derivatives (center panel) and first and second derivatives (right panel).}
    \label{figscgrMind}
\end{figure}

We have fitted a cubic b-spline basis to apply our methodology, as in the simulated schemes.
As reflected in Figure \ref{figscgrf}, the shapes of the two groups are different, and when applying the hypograph, epigraph (Figure \ref{figscgrind}) and its modified version (Figure \ref{figscgrMind}), the two groups have different behaviours despite that the obtained values seems to be overlapping one to another.

The best result in Table \ref{tablesgr} with the proposed methodology is obtained when applying k-means to the modified epigraph index to first and second derivatives with the Euclidean distance. The resultant clustering partition correctly classifies all boys, but fails to classify 3 girls as boys. The partition is very accurate. 

When applying functional k-means procedure (Table \ref{tabmgr}), the greater Purity coefficient is equal to 0.850 when applying a big value of $\rho$, $\rho = 1e+08$. Besides, apart from obtaining better metrics coefficients, our methodology reach an execution time almost 400 times smaller than when considering functional k-means strategy. 

Furthermore, applying the test based k-means technique (Table \ref{tablezgr}), obtains the best result with k-means initialization obtaining a Purity coefficient of 0.817. In summary, our methodology obtains the best result in terms of the three different metrics and in terms of execution time.

\begin{table}[htbp]
\vspace{-1cm}
\hspace*{-2cm}
\centering
\parbox{10cm}{

\footnotesize
\scalebox{0.9}{
\begin{tabular}{lcccc}
\hline 
 & Purity & Fmeasure & RI & Time \\ 
  \hline  \rowcolor{gray!10}
kmeans.dd2.MEI & 0.967 & 0.937 & 0.936 & 0.00498\\
\rowcolor{magenta!10}
kmeans.dd2.MEI & 0.967 &  0.937 & 0.936 & 0.00498 \\  
spc.dd2.MEI & 0.967 & 0.937 & 0.936 & 0.05681 \\ \rowcolor{orange!20}
svc.dd2.MEI & 0.967 & 0.937 & 0.936 & 0.00497 \\
\rowcolor{yellow!20}
kkmeans.\_dd2.MEI & 0.892 & 0.806 & 0.806 & 0.0476 \\\rowcolor{green!10}
svc.\_dd2.MEI & 0.892 & 0.806 & 0.806 & 0.02595 \\
\rowcolor{yellow!20}
kkmeans.dd2.MEI & 0.860 & 0.766 &0.756 & 0.03498 \\\rowcolor{green!10}
svc.\_d2.MEI & 0.806 & 0.690 & 0.684 & 0.01892 \\ \rowcolor{cyan!10}
kkmeans.\_d2.MEI & 0.817 & 0.701 & 0.698 & 0.03030 \\ \rowcolor{cyan!10}
kkmeans.$\_$dd2.MEI & 0.731 &  0.622 & 0.602 & 0.08166 \\ \rowcolor{magenta!10}
kmeans.d.EIHI & 0.699 &  0.664 & 0.575 & 0.00498 \\
 \rowcolor{cyan!10}
kkmeans.d.EIHI &  0.698 &  0.663 & 0.574 & 0.05100 \\ \rowcolor{gray!10}
 kmeans.d.EIHI &  0.698 &  0.663 & 0.574 & 0.00510 \\ \rowcolor{green!10}
 svc.dd2.MEI & 0.698 & 0.576 & 0.574 & 0.04088 \\ \rowcolor{yellow!20}
 kkmeans.$\_$d2.MEI & 0.677 & 0.560 & 0.558 & 0.02989 \\ \rowcolor{orange!20}
 svc.$\_$.EIHI &  0.655 &  0.621 & 0.543 & 0.00598 \\ \rowcolor{orange!20}
svc.d.EIHI & 0.655 &  0.592 & 0.543 &  0.00498 \\ \rowcolor{orange!20}
svc.d2.EIHI & 0.655 & 0.568 & 0.543 &  0.00495 \\
 \rowcolor{cyan!10}
 kkmeans.d2.EIHI & 0.655 & 0.568 & 0.543 & 0.04249 \\ \rowcolor{orange!20}
 svc.$\_$d.MEI &  0.645 &  0.540 & 0.537 & 0.01097 \\ \rowcolor{gray!10}
kmeans.$\_$d.MEI &  0.645 & 0.540 & 0.537 & 0.00597 \\ \rowcolor{magenta!10}
kmeans.$\_$d.MEI & 0.645 & 0.540 & 0.537 & 0.00507\\
\rowcolor{magenta!10}
kmeans.$\_$dd2.MEI & 0.634 &  0.534 & 0.531 & 0.00566 \\  

       \end{tabular}}
   }
\parbox{9cm}{
\footnotesize
\scalebox{0.9}{
\begin{tabular}[t]{lcccc}
  \hline
 & Purity & Fmeasure & RI & Time \\ 
  \hline \rowcolor{gray!10} 
kmeans.$\_$dd2.MEI & 0.634 & 0.533 & 0.531 & 0.00499 \\ \rowcolor{yellow!20} 
  svc.$\_$dd2.MEI & 0.634 & 0.533 & 0.531 & 0.00496 \\ 
\rowcolor{green!10}
svc.d2.EIHI & 0.623 & 0.538 & 0.525 & 0.0358 \\ \rowcolor{yellow!20}
kkmeans.d2.EIHI & 0.623 & 0.532 & 0.525 & 0.03098 \\ \rowcolor{cyan!10}
kkmeans.$\_$d.MEI &  0.634 &  0.533 & 0.531 & 0.04421 \\ \rowcolor{magenta!10}
kmeans.d2.EIHI & 0.613 &  0.598 & 0.520 & 0.00517 \\ \rowcolor{gray!10}
kmeans.d2.EIHI & 0.612 & 0.597 & 0.520 & 0.00598 \\
\rowcolor{yellow!20}
kkmeans.d.EIHI &  0.612 & 0.546 & 0.520 & 0.06006 \\ \rowcolor{yellow!20}
kkmeans.$\_$d.MEI &  0.602 &  0.518 & 0.515 & 0.03430 \\ \rowcolor{magenta!10}
kmeans.$\_$.EIHI & 0.591 &  0.590 & 0.512 & 0.00498 \\
  \rowcolor{gray!10}
kmeans.$\_$.EIHI &  0.591 &  0.590 & 0.511 & 0.00479 \\ \rowcolor{green!10}
svc.$\_$d.MEI &  0.580 &  0.508 & 0.507 & 0.02395 \\ \rowcolor{green!10}
svc.d.EIHI &  0.580 &  0.538 & 0.504 & 0.03291 \\
\rowcolor{cyan!10}
kkmeans.$\_$.EIHI &  0.580 & 0.582 & 0.501 & 0.04379 \\ \rowcolor{green!10}
svc.$\_$.EIHI &  0.580 &  0.503 & 0.501 & 0.04384 \\
spc.d.EIHI & 0.580  & 0.647 & 0.497 & 0.12768 \\ \rowcolor{gray!10} 
kmeans.$\_$d2.MEI & 0.581 &  0.497 & 0.496 & 0.00599 \\ \rowcolor{cyan!10}
kkmeans.dd2.MEI & 0.580 & 0.497 & 0.496 & 0.0249
\\ \rowcolor{magenta!10}
kmeans.$\_$d2.MEI & 0.580 &  0.497 & 0.496 & 0.00538 \\ \rowcolor{orange!20}
svc.$\_$d2.MEI & 0.580 &  0.497 & 0.496 & 0.00701 \\
spc.$\_$.EIHI & 0.580  & 0.620 & 0.495 &  0.08996 \\
spc.$\_$d2.MEI & 0.580 &  0.620 & 0.495 &  0.04521\\
spc.$\_$d.MEI & 0.580 &  0.630 & 0.494 & 0.0498 \\
   \hline
\end{tabular}}
}\hspace*{-1cm}
\caption{Mean results for growth data set considering Euclidean distance (gray), Mahalanobis distance (pink), a gaussian kernel (yellow), a polynomial kernel (blue), kernel k-means for initialization (green) and k-means for initialization (orange).}
\label{tablesgr}
\end{table}

\begin{table}[ht]
\centering
\footnotesize
 \begin{tabular}{lcccc}
  \hline
 & Purity & Fmeasure & RI & Time \\ 
  \hline
  $d\rho$, $\rho=1e+08$ & 0.850 & 0.747 & 0.742 & 1.93641 \\ 
  $d\rho$, $\rho=100$ & 0.753 & 0.634 & 0.624 & 1.74462\\ 
  $L^2$ & 0.656 & 0.551 & 0.544 & 0.50927\\ 
  $d\rho$, $\rho=0.001$ & 0.624 & 0.529 & 0.525 & 1.67513 \\ 
  $dk$, $k=2$ & 0.591 & 0.513 & 0.511 & 0.68651\\ 
  $d\rho$, $\rho=0.02$ & 0.581 & 0.517 & 0.508 & 2.10050\\ 
  $dk$, $k=3$ &  0.5806 &  0.4988 & 0.4974 & 0.4418 \\
  $d\rho$, $\rho=1$ & 0.581 & 0.496 & 0.495 & 2.18262\\ 
   \hline 
\end{tabular}
 \caption{Mean values of Purity, F-measure, Rand Index and execution time for the functional $k$-means procedure with truncated Mahalanobis distance, generalized Mahalanobis distance and $L^2$ distance to simulated data from growth data set.}
    \label{tabmgr}
\end{table}

\begin{table}[ht]
\centering
\footnotesize
\begin{tabular}{lcccc}
  \hline
 & Purity & Fmeasure & RI & Time \\ 
  \hline 
  kmeans  & 0.817 & 0.702 & 0.698 & 0.09893 \\
  kmeans $$++$$ & 0.666 & 0.552 & 0.5526 & 0.34084\\
  hclust & 0.666 & 0.552 & 0.525 & 0.09557\\ 
  random & 0.666 & 0.552 & 0.525 & 0.16951\\ 
   \hline
\end{tabular}
 \caption{Mean values of Purity, F-measure, Rand Index and execution time for the test based $k$-means procedure with four different initialization to simulated data from growth data set.}
\label{tablezgr}
\end{table}

\FloatBarrier
\subsection{Case study: Canadian weather data set}

Another popular real data set in the FDA literature, also included in \citet{ramsay} and in the `fda' R-package, which is the Canadian weather data set. This data set contains the daily temperature from 1960 to 1994 at 35 different canadian weather stations grouped in 4 different regions: Artic (3), Atlantic (15), Continental (12) and Pacific (5).

\begin{figure}[ht]
    \centering
    \includegraphics[scale=0.55]{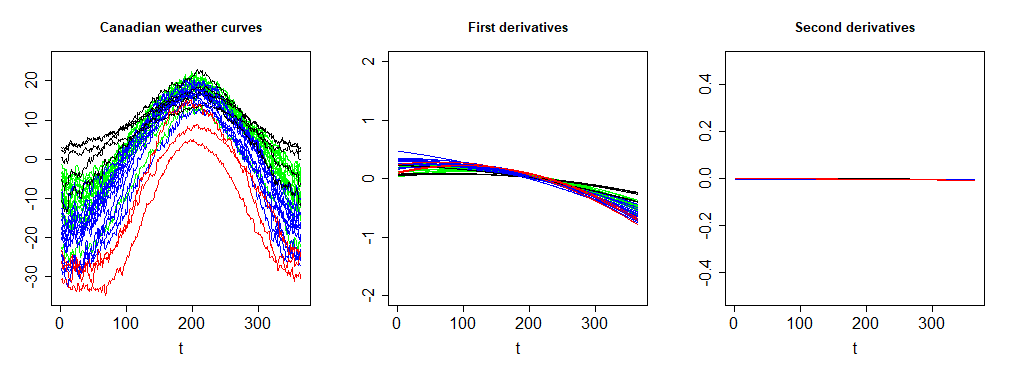}
    \caption{Canadian weather curves. Original data, first and second derivatives curves.}
    \label{figsccanf}
\end{figure}

\begin{figure}[ht]
    \centering
    \includegraphics[scale=0.5]{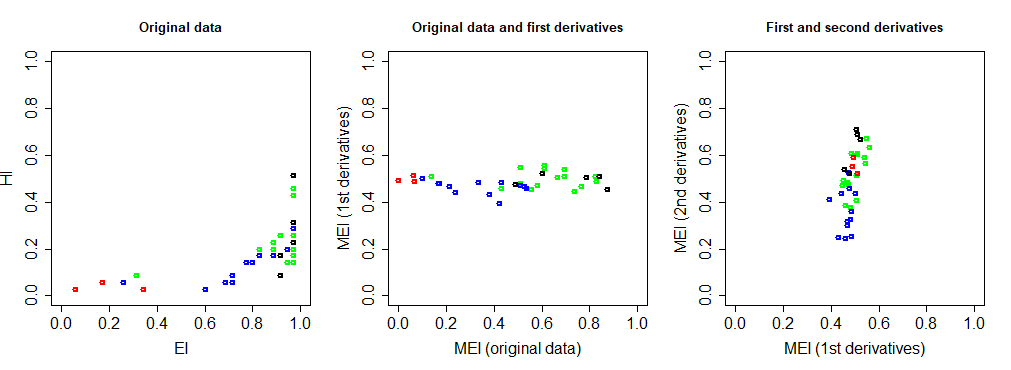}
    \caption{Canadian weather curves. Epigraph and hypograph index on the original data (left panel), the generalized epigraph index on the original data and first derivatives (center panel) and the generalized epigraph index on the first and second derivatives (right panel).}
    \label{figsccanind}
\end{figure}

We have applied a cubic B-spline basis to this data set, and then we have applied our methodology to it. First of all, we notice in Figure \ref{figsccanf} that first and second derivatives by themselves do not give much more information. Nevertheless, when applying the indexes and considering them together in Figure \ref{figsccanind}, they are able to distinguish in a better way between groups.

The best configuration between clustering method, indexes and variables is support vector clustering initialized with kernel k-means applied on the modified epigraph index of the original data and its second derivatives. The obtained Rand index is 0.719, while the F-measure is a smaller value equal to 0.510. This means that while the final configurations of groups are accurate, some groups are better classified than the others. For example, Pacific area obtains 6 elements inside when the correct number is 5, at the beginning it seems to be a good classification, but only two elements are correctly classified (Table \ref{confm1}). In general this classification obtains close results to the real ones. (ET=0.10655, Table \ref{tablescan}).

When considering functional k-means procedure (Table \ref{tabmcan}), the best result is obtained with truncated Mahalanobis distance (RI=0.784, F-measure=0.613). When talking about test based k-means (Table \ref{tablezcan}), the best result is obtained with a hierarchical clustering initialization (RI=0.764, F-measure=0.613).

Concerning execution time, these two alternative are much more time consuming than the proposed methodology, being this one 13 times faster than the functional k-means' best result, and 2 times better than test based k-means.

\begin{table}[htbp]
\hspace*{-2cm}
\centering
\parbox{10cm}{

\footnotesize 
\scalebox{0.9}{
\begin{tabular}{lcccc}
\hline 
 & Purity & Fmeasure & RI & Time \\ 
  \hline \rowcolor{green!10}
  svc.dd2.MEI & 0.771 & 0.513 & 0.722 & 0.05883 \\ \rowcolor{green!10}
svc.$\_$d2.MEI & 0.714 & 0.510 & 0.719 & 0.10655 \\ \rowcolor{cyan!10}
  kkmeans.$\_$d2.MEI & 0.686 & 0.511 & 0.714 & 0.02992\\ 
  spc.$\_$d.MEI & 0.657 & 0.421 & 0.686 & 0.03092\\ 
  spc.$\_$d2.MEI & 0.657 & 0.504 & 0.686 & 0.03191\\ \rowcolor{cyan!10}
  kkmeans.$\_$.EIHI & 0.686 & 0.450 & 0.684 & 0.03590 \\ \rowcolor{magenta!10}
  kmeans.$\_$d.MEI & 0.657 & 0.414 & 0.681 & 0.01096 \\ \rowcolor{magenta!10}
  kmeans.$\_$d2.MEI & 0.657 & 0.414 & 0.681 & 0.01193\\ \rowcolor{orange!20}
  svc.$\_$d.MEI & 0.657 & 0.414 & 0.681 & 0.00997 \\  \rowcolor{cyan!10}
  kkmeans.$\_$d.MEI & 0.657 & 0.406 & 0.671 & 0.02493\\ \rowcolor{green!10}
  svc.$\_$d.MEI & 0.657 & 0.391 & 0.671 & 0.07280\\ \rowcolor{yellow!20}
  kkmeans.$\_$d2.MEI & 0.629 & 0.393 & 0.667 & 0.05986 \\ \rowcolor{green!10}
  svc.d2.EIHI & 0.657 & 0.402 & 0.665 & 0.12418\\ \rowcolor{magenta!10}
  kmeans.d2.EIHI & 0.629 & 0.374 & 0.662 & 0.01200 \\ \rowcolor{magenta!10}
  kmeans.dd2.MEI & 0.629 & 0.385 & 0.662 & 0.01096\\ \rowcolor{yellow!20}
  kkmeans.dd2.MEI & 0.629 & 0.385 & 0.662 & 0.09314\\ 
       \end{tabular}}
   }
\parbox{9cm}{
\footnotesize
\scalebox{0.9}{
\begin{tabular}[t]{lcccc}
  \hline
 & Purity & Fmeasure & RI & Time \\ 
  \hline \rowcolor{orange!20} 
    svc.dd2.MEI & 0.629 & 0.385 & 0.662 & 0.01398\\
  spc.d2.EIHI & 0.629 & 0.378 & 0.652 & 0.03010 \\  \rowcolor{gray!10}
  kmeans.$\_$d.MEI & 0.629 & 0.373 & 0.650 & 0.01000 \\ \rowcolor{cyan!10}
  kkmeans.dd2.MEI & 0.571 & 0.355 & 0.645 & 0.06485 \\ \rowcolor{orange!20}
  svc.$\_$d2.MEI & 0.600 & 0.351 & 0.645 & 0.01196\\ \rowcolor{yellow!20}
  kkmeans.d2.EIHI & 0.600 & 0.359 & 0.640 & 0.03899 \\ \rowcolor{cyan!10}
  kkmeans.d2.EIHI & 0.543 & 0.336 & 0.635 & 0.03191\\ \rowcolor{gray!10}
  kmeans.$\_$.EIHI & 0.657 & 0.436 & 0.630 & 0.01196 \\ \rowcolor{magenta!10}
  kmeans.$\_$.EIHI & 0.657 & 0.436 & 0.630 & 0.00900\\ \rowcolor{orange!20}
  svc.$\_$.EIHI & 0.657 & 0.436 & 0.630 & 0.02136\\ 
  spc.$\_$.EIHI & 0.657 & 0.493 & 0.627 & 0.03967 \\ \rowcolor{yellow!20}
  kkmeans.$\_$.EIHI & 0.543 & 0.304 & 0.624 & 0.06072\\ 
  spc.dd2.MEI & 0.629 & 0.388 & 0.624 & 0.04587 \\ \rowcolor{orange!20}
  svc.d2.EIHI & 0.600 & 0.358 & 0.615 & 0.01097 \\ \rowcolor{yellow!20}
  kkmeans.$\_$d.MEI & 0.486 & 0.286 & 0.613 & 0.08537 \\ \rowcolor{green!10}
  svc.$\_$.EIHI & 0.486 & 0.279 & 0.600 & 0.04689\\ 
   \hline
\end{tabular}}
}\hspace*{-1cm}
\caption{Mean results for canadian weather data set considering Euclidean distance (gray), Mahalanobis distance (pink), a gaussian kernel (yellow), a polynomial kernel (blue), kernel k-means for initialization (green) and k-means for initialization (orange).}
\label{tablescan}
\end{table}

\begin{table}[ht]
\centering
\footnotesize
\begin{tabular}{|c||cccc||c|}
  \hline
 & Artic & Atlantic & Continental & Pacific & Total\\ 
  \hline \hline
  Artic &       3 &     0 &     0 &     0 & 3\\ 
  Atlantic &    1 &     8 &     2 &     4 & 15 \\ 
  Continental & 1 &     1 &    10 &     0 & 12\\ 
  Pacific &     0 &     3 &     0 &     2 & 5\\
     \hline \hline
  Total    &     5 &    12 &     12 &    6 & 35 \\
  \hline
\end{tabular}
\caption{Confusion matrix obtained from comparing real classification to the obtained with our proposal.}
\label{confm1}
\end{table}

\begin{table}[ht]
\centering
\footnotesize
 \begin{tabular}{lcccc}
  \hline
 & Purity & Fmeasure & RI & Time \\ 
  \hline
  $dk$, $k=3$ &  0.771 &  0.634 & 0.784 & 0.7937 \\
  $d\rho$, $\rho=1$ & 0.743 & 0.598 & 0.770 & 0.7697\\ 
  $d\rho$, $\rho=100$ & 0.743 & 0.552 & 0.746 & 0.7462\\ 
  $dk$, $k=2$ &  0.686 & 0.503 & 0.694 & 0.6941\\
  $d\rho$, $\rho=0.001$ & 0.686 & 0.489 & 0.681  & 0.6807 \\
  $d\rho$, $\rho=1e+08$ & 0.657 & 0.424 & 0.681 & 0.6807 \\
  $L^2$ & 0.686 & 0.489 & 0.681 & 0.6807\\
  $d\rho$, $\rho=0.02$ & 0.657 & 0.473 & 0.671 & 0.6706\\ 
 
   \hline 
\end{tabular}
 \caption{Mean values of Purity, F-measure, Rand Index and execution time for the functional $k$-means procedure with truncated Mahalanobis distance, generalized Mahalanobis distance and $L^2$ distance to simulated data from canadian weather data set.}
    \label{tabmcan}
\end{table}

\begin{table}[ht]
\centering
\footnotesize
\begin{tabular}{lcccc}
  \hline
 & Purity & Fmeasure & RI & Time \\ 
  \hline 
  hclust & 0.771 & 0.613 & 0.764 & 0.12433\\ 
  kmeans  & 0.714 & 0.532 & 0.731 & 0.08789 \\
  kmeans $$++$$ & 0.685 & 0.508 & 0.717 & 0.36513\\
  random & 0.600 & 0.427 & 0.657 & 0.12229\\ 
   \hline
\end{tabular}
 \caption{Mean values of Purity, F-measure, Rand Index and execution time for the test based $k$-means procedure with four different initialization to simulated data from canadian weather data set.}
\label{tablezcan}
\end{table}


\section{Choosing the number of clusters}
\label{sec:nclus}

In Sections \ref{sec:sim} and \ref{sec:realdata}, the number of clusters was set in advance. Nevertheless, choosing the correct number of clusters before applying a clustering technique is a challenge. In \citet{martino} and \citet{zambom2019} they both fix the number of clusters before performing the classification.

To overcome this problem, we have considered the Silhouette index. Let $x_i$ be one of the considered points, and let $a(x_i)$ be the average distance of $x_i$ with respect to all other points in its cluster and $b(x_i)$ be the lowest average distance of $x_i$ to any other cluster of which $x_i$ is not a member. Then $$s(x_i) = \frac{b(x_i)-a(x_i)}{\max\{a(x_i), b(x_i)\}}.$$ The silhouette index ranges from -1 to 1 where a positive value means that the object is well matched to its own cluster, and a negative value means that the object is bad matched to its own cluster. The average silhouette $$\bar s = \frac{1}{n} \sum_{i=1}^n s(x_i),$$ gives a global measure of the election of the clusters, such that the more positive, the better the configuration. Thus, we choose the number of clusters as the one providing the greater average silhouette.

We simulate 100 times each of the scenarios that have been considered in Section \ref{sec:sim} and apply the mean silhouette to obtain the optimal number of clusters. In Table \ref{table1} appears the number of times corresponding to each possible number of cluster. In this case we have consider numbers between 2 and 6, but the list could be any other. When the real number of clusters is two, the correct number of clusters is always obtained. Thus, we can conclude that this procedure works well for two clusters. However, this strategy is not consistent with three clusters and it is necessary to look for an alternative.

\begin{table}[htb]
\begin{center}
\footnotesize
 \begin{tabular}{c c c c c c} 
 \hline
  & 2 & 3 & 4 & 5 & 6 \\ [0.5ex] 
 \hline
 \textit{S 1-2} & \textbf{99} & 0 & 0 & 0 & 1 \\
 \textit{S 1-3} & \textbf{100} & 0 & 0 & 0 & 0 \\
 \textit{S 1-4} & \textbf{100} & 0 & 0 & 0 & 0 \\
 \textit{S 1-5} & \textbf{62} & 5 & 5 & 7 & 21 \\
 \textit{S 1-6} & \textbf{72} & 5 & 4 & 6 & 13 \\
 \textit{S 1-7} & \textbf{89} & 2 & 1 & 3 & 5 \\
 \textit{S 1-8} & \textbf{100} & 0 & 0 & 0 & 0 \\
 \textit{S 1-9} & \textbf{100} & 0 & 0 & 0 & 0 \\
  \textit{S 10-11} & \textbf{71} & 15 & 8 & 4 & 2 \\
 \textit{S 10-12} & \textbf{73} & 15 & 5 & 3 & 4 \\
  \textit{S 13-14-15} & 84 & \textbf{16} & 0 & 0 & 0 \\
 \textit{S 16-17-18} & 100 & \textbf{0} & 0 & 0 & 0 \\
 \textit{S 19-20-21} & 100 & \textbf{0} & 0 & 0 & 0 \\
 \hline
\end{tabular}
\end{center}
\caption{Distribution of the number of clusters suggested when applying Silhouette for each Scenario simulated 100 times.}
\label{table1}
\end{table}

We have also considered 30 different indexes for multivariate clustering available in the R package `NbClust' and which are fully explained in \citet{nbclust}. With all these indexes, we have not find consistent techniques for two and three clusters. Thus, this is a still open research line.

\section{Discussion}
\label{sec:conc}

In this paper, we propose a new methodology for clustering functional data that is competitive with respect to the existing ones, and that is significantly better in terms of execution time. Our methodology is based on converting a functional problem into a multivariate problem through the use of the epigraph, the hypograph indexes, their generalized versions and multivariate clustering techniques. It has been compared to two recent procedures for clustering functional data, outperforming them in most of the cases and in all cases when concerning execution time. Finally, the code needed to carry out this analysis and to apply our technique is available in the GitHub repository: \url{https://github.com/bpulidob/Functional-clustering-via-multivariate-clustering}.

In the new proposal, we have set the number of clusters in advance. Despite that, an strategy for choosing the number of clusters has been tried in Section \ref{sec:nclus}, without obtaining a consistent technique for two and three clusters. Thus, setting the number of clusters prior applying the clustering technique is a question still open for further research.






\renewcommand\bibname{References}
\addcontentsline{toc}{chapter}{References}
\printbibliography

\end{document}